\documentclass[galaxies,article,accept,pdftex,moreauthors]{mdpi} 
\usepackage{diagbox}

\DeclareMathOperator{\sech}{sech}
\newcommand{\X}{\mathrm{X}}
\firstpage{1} 
\pubvolume{1}
\issuenum{1}
\articlenumber{0}
\pubyear{2026}
\copyrightyear{2025}
\datereceived{2 December 2025} 
\daterevised{26 December 2025} 
\dateaccepted{29 December 2025} 
\datepublished{ } 
\hreflink{https://doi.org/} 

\Title{Boxy/Peanut 
 Bulges: Comparative Analysis of EGIPS Galaxies and TNG50 Models}

\Author{Anton 
 Smirnov $^{1,}$*\orcidA{}, 
 Alexander Marchuk $^{1,2}$, Viktor Zozulia $^{1,2}$, Natalia Sotnikova $^{1,2}$ and Sergey Savchenko $^{1,2}$}

\AuthorNames{Anton Smirnov, Alexander Marchuk, Viktor Zozulia, Natalia Sotnikova and Sergey Savchenko}

\address{%
$^{1}$ \quad Central (Pulkovo) Astronomical Observatory of 
 the Russian Academy of Sciences, 
 Pulkovskoye Chaussee 65/1, \mbox{196140 St. Petersburg, Russia}; a.a.marchuk+astro@gmail.com (A.M.); vdzozulia.astro@gmail.com (V.Z.); n.sotnikova@spbu.ru (N.S.); s.s.savchenko@spbu.ru (S.S.) 
\\
$^{2}$ \quad St. Petersburg State University, 
 Universitetskij pr.~28, 198504 St. Petersburg, Russia 
}

\corres{Correspondence: zeleniikot@gmail.com}

\abstract{
We investigated the properties of boxy/peanut-shaped (B/PS)
    bulges in a sample of \mbox{71 galaxies} from the Edge-on Galaxies
    in the Pan-STARRS Survey (EGIPS)
    and \mbox{20 simulated galaxies} from Illustris TNG50 using multicomponent
    photometric decomposition. For each real and simulated galaxy,
    we obtained a suitable photometric model in which the B/PS bulge was
    represented by a dedicated 2D photometric function.
    For real galaxies, we found that more flattened X-structures are
    generally residing in larger B/PS bulges.
    When tested against the galaxy masses, we verified that
    both larger bulges and more flattened X-structures are typically
    found in more massive galaxies. Since large bars are also known to
    reside in more massive galaxies, we conclude that the
    flatness of X-structures in larger B/PS bulges has a physical origin,
    rather than being solely a result of projection effects due to differences
    in observed bar viewing angles.
    When comparing the properties of B/PS bulges
    between EGIPS galaxies and TNG50 galaxies, with bars rotated
    for different viewing angles, we found that B/PS bulges
    in TNG50 are considerably smaller and less luminous in terms of total intensity.
    This is consistent with previous studies of bar properties in TNG50,
    indicating the B/PS bulges in TNG50 differ from those in real galaxies,
    as do their \mbox{parent bars}.
}

\begin{document}
\section{Introduction}

Boxy and peanut-shaped (B/PS) bulges are a distinct type of stellar subsystem frequently observed in the central
regions of edge-on disc galaxies, distinguished by their morphology and dynamical properties~\cite{Shaw1987, deSouza_DosAnjos1987, Lutticke_etal2000,ED13,Yoshino_Yamauchi2015,Erwin_Debattista2017, Marchuk2022}. These bulges are believed to be intrinsically linked to the bars, being the extension of the latter in the direction perpendicular to the disc plane. This implies that the ``building blocks'' of galaxies, namely the 3D regular stellar orbits of stars, constituting bars viewed face-on and boxy/peanuts viewed edge-on are essentially the same~\cite{Pfenniger_Friedli1991,Skokos_etal2002}. The fact that the growing bars are usually not confined to the disk plane has been proved in numerous simulation studies~\cite{Combes_Sanders1981,Raha_etal1991,Athanassoula_Misiriotis2002, Oneil_Dubinski2003, Athanassoula2005, MartinezValpuesta_etal2006, Debattista_etal2006, Wozniak_Michel-Dansac2009, Saha_Gerhard2013, Fragkoudi_etal2017, Smirnov_Sotnikova2018,Sellwood_Gerhard2020}. Conversely, it is commonly assumed that every observed boxy/peanut-shaped bulge originates from a corresponding bar. Direct evidence supporting this has been obtained for a sample of galaxies with B/PS bulges through kinematical analyses, which show that these structures exhibit cylindrical rotation~\cite{Bureau_Freeman1999, Chung_Bureau2004}. Furthermore, when a stellar disk is viewed at slight or moderate inclination, both the boxy/peanut features and the planar bar can be simultaneously observed~\cite{ED13, Kruk_etal2019, 2021A&A...647A..20P}.


This study primarily focuses on the observational properties of boxy/peanuts. Consequently, the theoretical developments in the field, including orbital studies and studies of B/PS bulges formation mechanisms, are not discussed in extensive detail. Instead, we limit ourselves to highlighting a few key facts that are also significant from an \mbox{observational standpoint}.

First, the bulge morphology, whether boxy or peanut-shaped, is largely determined by the so-called bar viewing angle, an angle between the line of sight (LoS) and the bar major axis. In general, the closer the bar major axis is aligned with the LoS (end-on view), the more boxy the bulge appears, whereas the most pronounced peanut morphology is observed when the bar major axis is oriented perpendicular to the LoS (side-on view).

A second note concerns terminology. Above the disk plane the, B/PS bulges frequently exhibit distinctive intensity
enhancements  that have an X-like shape. Those are usually referred  as ``X-structures'' and sometimes understood as
a separate entity from underlying B/PS bulge. However, the orbital studies tell
us that no orbits constitute the B/PS bulge and X-structures separately, 
 but the latter make their
appearance due to nature of orbits in the B/PS bulge~\cite{Patsis_Katsanikas2014a,Portail_etal2015b,Parul_etal2020}. The stars moving along 3D orbits in B/PS bulge prefer to stay longer times at some specific points, typically furthest from the center and highest from the disk plane. The X-structure rays are formed by alignment of such points for orbits with different extensions of major axes in physical space.
\par
From an observational perspective, B/PS bulges are rather complicated objects to study.
This complexity primarily stems from the fact that they are observed in the central regions
of edge-on or nearly edge-on galaxies, where the line-of-sight (LoS) intensity represents
the superposition of multiple structural components, e.g., the B/PS itself
and the stellar disk. Additional structural features, including other types of bulges and rings,
may also be present. It is well-known that, except B/PS bulges,
central parts of galaxies can host two other types of bulges
~\cite{2004ARA&A..42..603K,2B_Kormendy, 2B_Erwin}:
classical ones, usually viewed as the remnants of minor galaxy merges,
and the so-called pseudobulges, forming through the bar-induced gas inflow
to the galaxy central regions. Because different bulge types can
coexist within the same galaxy, interpreting observational
results can be challenging, as one may entirely miss to which physical
component observable properties such as bulge-to-total ratio ($B/T$),
scale lengths, etc., should be attributed. Nevertheless, boxy/peanuts
are important objects to study since they provide a unique possibility
to study bars vertical structure, the properties of which should
be somehow connected with conditions in the stellar disk, where the bar \mbox{is formed}.

\textcolor{black}{Observational studies of B/PS bulges are vast in scope, ranging from
statistical \mbox{studies~\cite{Shaw1987,deSouza_DosAnjos1987,Lutticke_etal2000,ED13,
    Yoshino_Yamauchi2015,Erwin_Debattista2017}}, which provide important information
on the abundance of B/PS bulges in the local Universe and beyond~\cite{Kruk_etal2019},
    to more sophisticated analyses,
including the isophotal analysis~\cite{ED13, CG16},
    structural analysis via unsharp-masking~\cite{LS17},
and photometric \mbox{decompoisition~\cite{Savchenko_etal2017, Xold}}.
So far, it has been found that B/PS are observed in nearly 20--40\% of disk galaxies,
    and the frequency strongly depends on the host galaxy mass, showing
    a sharp increase at 
 $M\gtrsim10^{10.3-10.4}M_\odot$~\cite{Erwin_Debattista2017,Marchuk2022}.
}

\textcolor{black}{
    As for the structural parameters, B/PS bulges typically
    exhibit a substantial contribution to
    the total galaxy luminosity, quantified by the bulge-to-total ratio $B/T$,
    with median values for different samples ranging from about 0.2 to 0.5~\cite{Marchuk2022}
   (keep in mind that the $B/T$ values referred to here are obtained
    by means of automatic bulge plus disk
    decomposition in~\cite{sdss2011, EGIS, Yoshino_Yamauchi2015}
    and are likely to overstate the contribution of the B/PS bulge itself.). 
    Comparing $B/T$ values between galaxies with and without X-shaped features,
    ref. 
 \cite{Marchuk2022} also found that the typical difference
    $\Delta \left[(B/T)_\X-(B/T)_{no\,X}\right]\sim0.1-0.2$ depending on the sample considered.
    For B/PS bulge sizes, refs.~\cite{ED13,LS17} found that boxy/peanut features
typically extended up to about 0.2-0.5 of the bar size, with a median value around 0.4.
Refs. \cite{Savchenko_etal2017,Marchuk2022} measured
    the lengths of X-structure rays and found that they typically extend
    to about 1.2 disk scale lengths.
}

\textcolor{black}{
    A distinctive characteristic of B/PS bulges,
    the angle between the X-structure rays and the disk plane,
has also been measured in various studies.
Using a sample of numerical models, ref. \cite{Smirnov_Sotnikova2018}
    showed that this angle evolves with time,
    decreasing as the bar grows. Its value also depends on the initial conditions
    in the disk and the dark matter halo, making it a
    valuable metric of a galaxy’s secular evolution.
    Observational studies~\cite{LS17, Savchenko_etal2017, Xold} find
    that X-structure angles range from $25^\circ$ to $45^\circ$
    (the measurements from \cite{LS17} can be converted to angles
    by taking the arctangent of $b/a$ provided there),
    which is consistent with numerical results of~\cite{Smirnov_Sotnikova2018},
    when the bar viewing angle is taken into account~\cite{Xold}.
}

The primary focus of the present work is to expand our
knowledge about the statistics of parameters of B/PS bulges
(luminosity fraction, sizes, shapes, X-structure angles) and
to compare those with what one obtains in up-to-date cosmological
simulations like TNG50 for cosmological models. Compared to previous studies,
where the B/PS bulges were either analyzed under the common roof of ``pseudobulges''
or only some geometric properties of such bulges were measured,
here we want to analyze specially boxy/peanuts by means of 2D photometric decomposition,
while separating the contribution
of the B/PS bulges from other bulges (if present).
\textcolor{black}{This is in line with the studies of~\cite{2B_Kormendy, 2B_Erwin},
which showed that bulges in disk galaxies can be composite structures,
hosting multiple types of bulges and/or nuclear components, all of which
contribute to the central intensity.}
This work can be considered a continuation of our previous
observational studies of B/PS bulges,
\mbox{specifically~\cite{Savchenko_etal2017, Xold, Marchuk2022}}, where B/PS bulges
of real galaxies were analyzed through photometric decomposition.
In the present study, we focus more on the sizes and contributions
of B/PS bulges to the total intensity.
Additionally, we perform a more sophisticated comparison between real
and simulated galaxies than in~\cite{Xold}, by converting the mass profiles
of simulated galaxies into intensity maps in specific bandwidths
using \mbox{SKIRT (v9.0)}~\cite{2011ApJS..196...22B,2015A&C.....9...20C}~simulations. 

\textcolor{black}{The present work is structured as follows.
In Section~\ref{sec:data}, we describe our EGIPS and
Illustris TNG50 datasets and outline the procedure used to
generate realistic images of TNG50 galaxies for different bar
viewing angles with SKIRT. Section~\ref{sec:decomposition}
presents an overview of our multicomponent photometric decomposition pipeline,
    in which the B/PS bulge is included as a separate component.
    In Section~\ref{sec:res}, we investigate the parameters of B/PS bulges
    obtained from the photometric decomposition and analyse their correlations
    with the properties of other structural components,
    such as the disk and additional bulges.
    Section~\ref{sec:discussion} discusses our results in the context
    of previous studies of B/PS bulges and bar/bulge properties in TNG50.
    Finally, Section~\ref{sec:conc} summarises our conclusions.}
\section{Data}
\label{sec:data}
\begin{figure}[H]
    \includegraphics[width=0.75\textwidth]{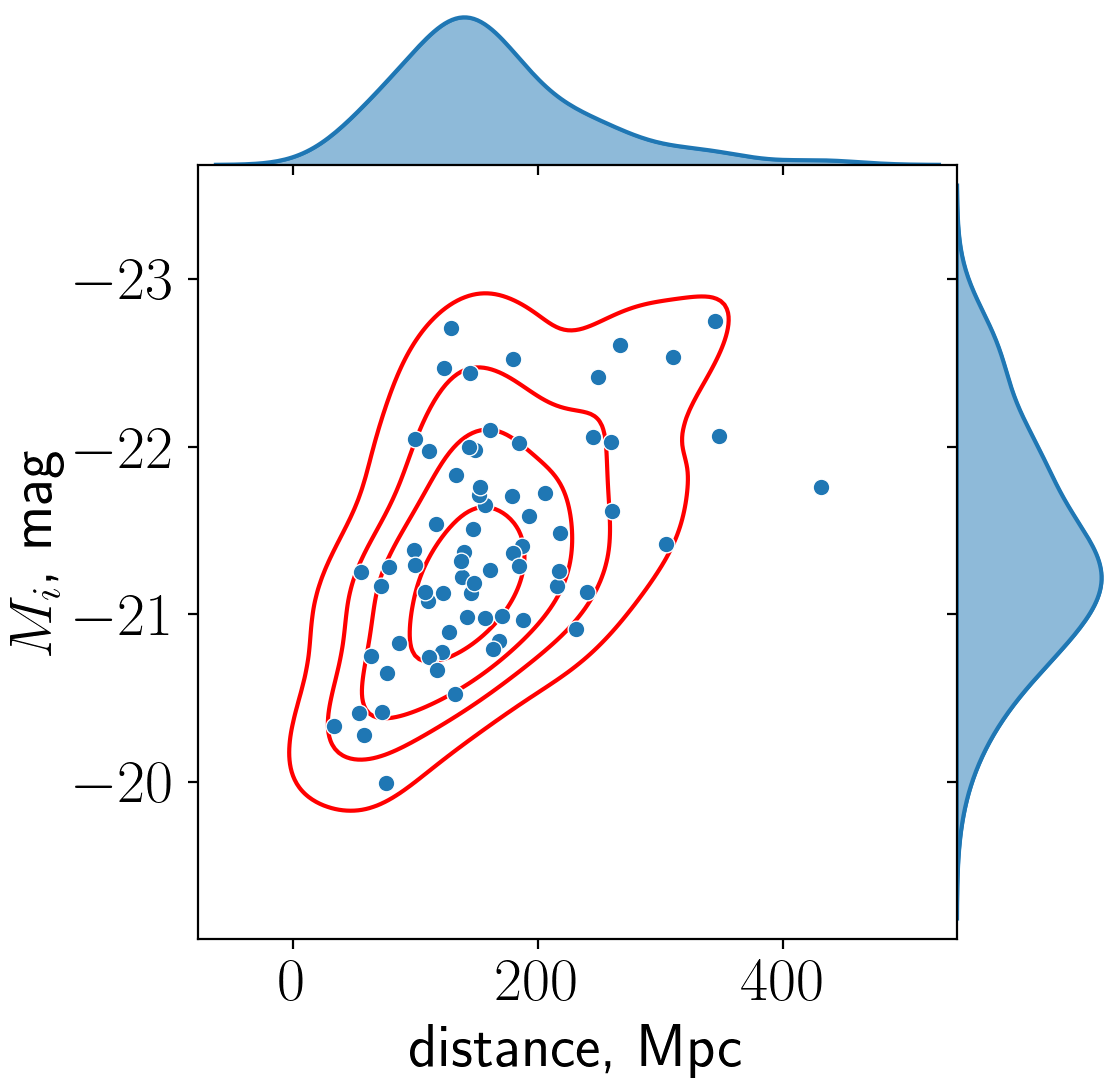}
    \caption{Redshift-derived 
 spectroscopic distance versus absolute stellar magnitude in the $i$ band for the sample of galaxies with boxy/peanut-shaped bulges, studied in the present work.}
    \label{fig:mag_dist}
\end{figure}

Two data sets were employed in the present study.
The first consists of a subsample of galaxies taken from the
Edge-on Galaxies in the Pan-STARRS survey (EGIPS,~\cite{2022MNRAS.511.3063M}).
The Pan-STARRS survey is conducted
with a 1.8 m telescope at Haleakalā Observatory (Hawaii)
and covers the Northern Hemisphere, as well as part of the Southern
Hemisphere down to Dec. = $-$30 deg. The data are available in
five broad bands $(g, r, i, z, y)$, with the resolution of
0.25 arcsec per pixel.
Candidate galaxies for the present analysis were selected based on the outcomes of an automatic photometric
decomposition conducted for each EGIPS galaxy in the $i$ band (Savchenko et al., 2026, in prep., see \url{https://www.sao.ru/edgeon/catalogs.php?cat=PS1cand1_imfit2},  
).
Specifically, galaxies hosting B/PS bulges were identified
by the characteristic X-shaped patterns in the residuals,
which were obtained by subtracting the photometric models
of the disk and bulge from the original image.
\textcolor{black}{A similar technique was employed in~\cite{Marchuk2022} to identify galaxies
with B/PS bulges in the DESI (Dark Energy Spectroscopic Instrument) Legcay survey~\cite{DESI}.}
Galaxies exhibiting such residues were conveniently identified
from the results of automatic decomposition, with a total number of 83 candidates.
Subsequently, 12 galaxies were further excluded based on photometric decomposition
results when the boxy or peanut-shaped features were deemed too weak,
leaving a final sample of 71 galaxies with prominent B/PS bulges.
Absolute magnitudes and spectroscopic distances
(relative to the Cosmic Microwave Background, CMB) of
selected galaxies are shown in Figure~\ref{fig:mag_dist}.
\textcolor{black}{Spectroscopic redshifts are compiled from the HyperLEDA (\url{http://atlas.obs-hp.fr/hyperleda/})~\cite{Hyperleda} and
the NASA/IPAC Extragalactic Database (\url{https://ned.ipac.caltech.edu/}). 
Absolute magnitudes are taken from the EGIPS catalogue
~\cite{2022MNRAS.511.3063M} and correspond to
Kron magnitudes corrected
for Galactic extinction~\cite{DustCorrection}.}
In general, our sample consists of relatively bright galaxies,
with a median absolute magnitude of $M_i=-21.3$ mag and a median distance of about 150 Mpc.

The second data set comprises cosmological Milky Way-like models, a sample compiled
from Illustris TNG50 simulations
by~\cite{2024MNRAS.535.1721P}. The sample consists of 198 relatively massive disk galaxies with stellar masses $M_*(r<\, $30 kpc$)=10^{10.5-11.2}M_\odot$ and no massive companions; specifically, they do not have massive satellites $M_\star > 10^{10.5} M_\odot$  within $r<500$ kpc from the galactic center. For a detailed description of corresponding simulations, interested readers are referred to the original papers by~\cite{10.1093/mnras/stz2338,2024MNRAS.535.1721P}. Here, we should  note that TNG50 is one of the most novel magnetohydrodynamic simulations, featuring the self-consistent evolution of dark matter, stars, gas, supermassive black holes, and magnetic fields, implemented through the AREPO code~\cite{2010MNRAS.401..791S}. All these components are evolved from redshift $z=127$ to $z=0$ within a cubic volume of $51.7$ Mpc (comoving units), with periodic boundaries. TNG50 benefits from a high spatial resolution, with galaxy models typically represented by about one million particles, and a softening length of around 300 pc at lower redshifts for star and dark matter particles. This resolution enables analysis of the vertical structure of galactic disks, as highlighted by~\cite{2024MNRAS.535.1721P}.
\par
Not all models from the sample of~\cite{2024MNRAS.535.1721P} host bars,
and naturally, not all of them exhibit B/PS features when viewed edge-on
(see also~\cite{Tng_bps}).
Therefore, the sample required additional processing.
To obtain a sample of models with B/PS bulges,
we prepared a preliminary set of images for each model,
generated directly from the stellar mass distribution,
featuring face-on and edge-on projections, which were then inspected visually.
First, we selected models hosting bars (154 out of 198). Then, we rejected models
that did not exhibit peanut features when viewed
edge-on with the bar rotated perpendicularly to the LoS and/or
the disk structure is significantly disturbed. This step cuts most of the sample by far,
leaving only 56 cosmological models. Some candidates were further
rejected after preliminary photometric decomposition,
if the B/PS bulge photometric model were trying to account for prominent
non-symmetric disk features, polar rings, etc.
Our resulting sample of TNG50 galaxies consists of 20 models with prominent B/PS bulges.
\par
\begin{figure}[H]
    \includegraphics[width=0.9\linewidth]{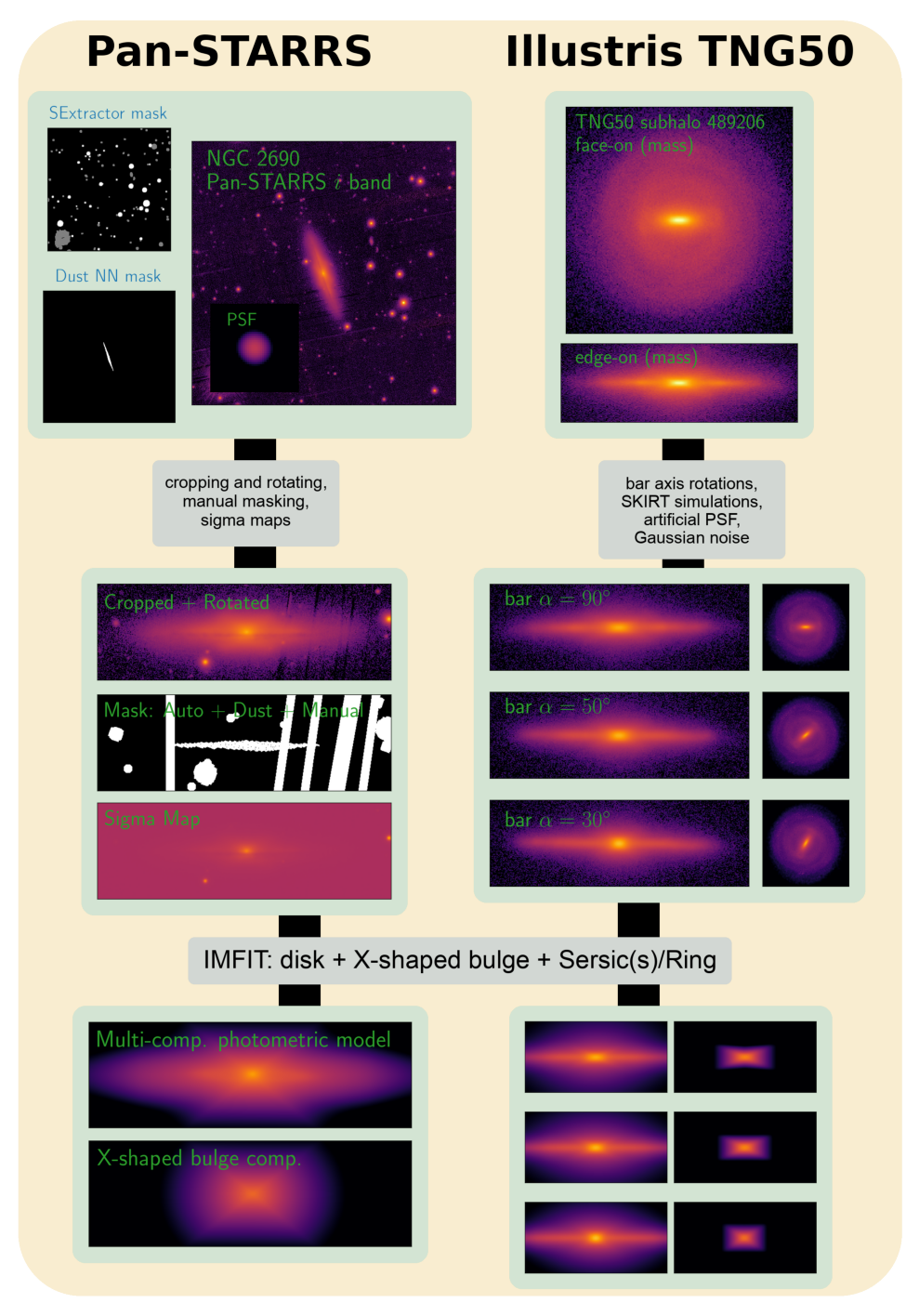}
    \caption{A 
 general scheme of the galaxy processing pipeline used in the present work for
    EGIPS galaxies (\textbf{left} 
 side) and Illustris TNG50 galaxies (\textbf{right} side).}
    \label{fig:pipeline}
\end{figure}

To obtain realistic images of simulated galaxies, we employed the SKIRT radiative transfer
code~\cite{2011ApJS..196...22B,2015A&C.....9...20C}. SKIRT is a versatile software tool capable of
transforming three-dimensional simulation datasets into FITS images at specified wavelengths
by incorporating the density distributions, softening lengths, and properties of stellar populations.
These data are available directly from the TNG50 snapshots. To enable comparison with observational data,
we prepared multiple images for each simulated galaxy in $i$ band featuring different
bar viewing angles, specifically, 90 deg, 50 deg, and 30 deg, to account for projection effects
(see Figure~\ref{fig:pipeline}).
Additional simulations parameters are as follows:

\begin{enumerate}
    \item TNG50 distances and masses are given in units of $h=H_0/100$, where $H_0$ is the
    Hubble constant. For example, masses are measured in units of
    $h^{-1} 10^{10} M_\odot$. Those quantities were converted to appropriate values by dividing them by $h=0.6774$.
    \item Each galaxy is simulated in SKIRT using $10^9$ photon packets.
    \item Each image encompasses a region extending from $-$22.5 to 22.5 kpc from the galaxy center in the disk plane,
    and from $-$7.5 to 7.5 kpc along the vertical direction.
    \item The images have a resolution of 300 $\times$ 100 pixels.
    The linear scale of each pixel is then 150 pc.
    Assuming a distance of 150 Mpc, this gives a pixel scale of 0.2 arcsec, which is slightly better than
    Pan-STARRS resolution.
    \item Each stellar particle is assumed to represent a single stellar population,
    characterized by a Chabrier initial mass function~\cite{2003PASP..115..763C},
    with defined formation time, metallicity, and the initial mass (mass at formation time).
    For the spectral energy
    distribution,
    we adopt the Bruzual and
    Charlot stellar population synthesis model~\cite{2003MNRAS.344.1000B}.
    \item As the smoothing length, we adopt the distance to
    the 32 nearest particles (see ~\cite{10.1093/mnras/stu2592}),
    a parameter that is directly available in the TNG50 simulation snapshots.
    However, as noted in the SKIRT description (see \url{https://skirt.ugent.be/root/_home.html}),
    if the softening length is uncapped, it can lead to the appearance of large artificial blobs,
    as some distant
    particles in a given simulation snapshot can have very large distances
(e.g., >100 kpc). Therefore, we capped the softening length at 1.5 kpc. In practice, we verified that the image does not change significantly as long as at least several
pixels are covered by the softening length. For our chosen softening value,
the softening covers 10 pixels.
    \item The resulting images were convolved with a
Gaussian kernel of FWHM 5 pixels (1~arcsec, assuming a distance of 150 Mpc),
simulating the effect of the Pan-STARRS PSF in the $i$-band.

\end{enumerate}


\section{Photometric Decomposition}
\label{sec:decomposition}
\subsection{Preprocessing}


\textcolor{black}{In this section we describe the image preparation for the photometric
decomposition process (see Figure~\ref{fig:pipeline}
for the illustration of the main steps). We start with the
image cropping to obtain a field centered on the galaxy that
fits the entire galaxy with margins of about the optical diameter of the galaxy.
Next, to exclude the background objects
from the analysis, we make a binary mask based on the objects catalogue
produced by the SEXTRACTOR
package~\citep{sextractor}
(the galaxy under consideration was excluded from the catalogue by its coordinates). When
necessary we manually modified the mask to include image artifacts or
extended features of oversaturated stars.  In
order to take into account the image smearing by the atmosphere and the telescope optics, we make PSF images.
To do so we utilized the
PSFEX software 
 package~\citep{psfex}. This package automatically extracts appropriate stars on a field based on
a SEXTRACTOR catalogue and combines them into an output PSF image.}

\textcolor{black}{Since the dust attenuation can significantly impact the results of the
decomposition~\citep{gadotti_2010}, especially
for highly inclined galaxies \citep{savchenko_2023},
we decided to mask regions close to a galactic disk plane, where
the dust impact is most severe. To produce a dust mask we used a
neural network model described in~\cite{savchenko_2023}.
This model has a U-Net architecture~\citep{ronneberger_2015} and
was trained such that it takes a
stack of $g$-, $r$-, and $i$-band images as an input and produces
    a binary mask that covers the regions most affected by
dust (see~\cite{savchenko_2023} for more details).
The dust masks obtained via the neural network were inspected visually and
    further modified manually if deemed necessary.
    Overall, the mask used during the decomposition process is a combined mask of
background objects and dusty regions.}

\textcolor{black}{For the illustrative purposes we rotated
the images to approximately align the disk plane with the $x$-axis.}

\subsection{General Setup}
To obtain photometric models for our selected galaxies from
EGIPS and synthetic images of cosmological models, we utilize
the IMFIT software 
 package (v1.8.0)~\citep{imfit}.
Each image is fitted using a set of predefined 2D or 3D functions (see below).
By default, IMFIT searches for the best-fit model
by exploring the parameter space using the Levenberg--Marquardt algorithm~\citep{LM}.
However, since our models typically consist of at least three components,
namely disk, B/PS bulge, and additional bulge and, thus, involve many parameters,
we first run the Nelder--Mead simplex algorithm~\citep{NM},
which is slower but less prone to becoming trapped in local minima.
After this initial step, we perform a more precise
search using the Levenberg--Marquardt algorithm.

The best-fit parameters are found by minimising  $\chi^2$ statistics: 
\begin{equation}
 -2 \ln \mathcal{L} =  \chi^2 = \sum_{i=0}^N w_i \, (I_{m,i}-I_{d,i})^2,
\label{eq:stat}
\end{equation}
where $\mathcal{L}$ is the likelihood of a given model,
$w_i$ are the pixel weights, and $I_{m,i}$
and $I_{d,i}$ are the intensities of the individual pixels in the
photometric model and in the observed image, respectively.
Weights $w_i$ are reverse squares of Gaussian errors $w_i = 1/\sigma_i^2$, where
\begin{equation}
    \sigma^2_i = (I_{d,i}/g + \sigma^2_\mathrm{back}).
    \label{eq:sigma_map}
\end{equation}
 $g$ 
 is the gain and $\sigma_\mathrm{back}$ is the standard deviation of the background,
measured over $10^4$ boxes of 10 arcsec size that are randomly
distributed across the Pan-STARRS field of a galaxy.
Note that errors in IMIFT are usually calculated
based on the specified readnoise $\sigma_\mathrm{rdn}$ value,
which we ignore here. This is because, in the course of this work, we
found that readnoise values stored in Pan-STARRS data for corresponding
galaxies do not accurately reflect the noise amplitude.
\textcolor{black}{For example, there are some cases where $\sigma_\mathrm{back}$
is two orders of magnitude greater
than $\sigma_\mathrm{rdn}$ provided in the Pan-STARRS corresponding fields.
We assume that the difference arises from the original background intensity,
which was  subsequently subtracted
from the images by the Pan-STARRS pipeline, and, unfortunately, is unavailable
for Pan-STARRS data. The use of $\sigma_\mathrm{back}$ in Equation~(\ref{eq:sigma_map}) prevents
the photometric model from fitting the noisy parts of the
images that are underestimated by $\sigma_\mathrm{rdn}$, while preserving
the Poisson noise associated with the original data counts.}

\textcolor{black}{For our TNG50 images produced by SKIRT, we introduced Gaussian noise with the amplitude
$\sigma=0.03$ MJy/sr corresponding to the median value of background noise $\sigma_\mathrm{back}$
measured for our EGIPS sample and use the same $\chi^2$ statistic described by Equation~(\ref{eq:stat}).
We do not add Poisson noise to TNG50 images, since it is difficult to
obtain proper photon counts estimate for Pan-STARRS
images (see \url{https://outerspace.stsci.edu/spaces/PANSTARRS/pages/298812205/PS1+FAQ+-+Frequently+asked+questions}).
}

\subsection{Photometric Functions}
\label{sec:phot_func}
In course of this work, we assume that every galaxy image can be decomposed into
a set of 2D functions, which we describe below. 

\textls[25]{The first one is the most commonly used in photometric studies,
the S\'{e}rsic \mbox{function~\cite{Sersic}:}}
\begin{equation}
I(a)=I_e \exp \left[-\nu_n\left(\left(\frac{a
}{r_e}\right)^{1/n}-1\right)\right],
\label{eq:ser}
\end{equation}
where $r_e$ is the effective radius (the radius containing half of the total flux),
$I_e$ is the brightness at $r_e$, $n$ is the S\'{e}rsic index, $\nu_n$ is a
function depending on $n$~\cite{Caon_etal1993}. Note that $a$ here is the elliptical radius,
determined from the ellipticity of the isophotes $\varepsilon$ and the pixel coordinates $(x, y)$ as
\begin{equation}
a (x,y,\varepsilon)=\sqrt{x^2 + y^2/(1-\varepsilon)^2}.
\end{equation}
We use S\'{e}rsic function primarily to describe an
additional central structure that sometimes appears on top of the B/PS bulge.
This component is usually smaller than the B/PS bulge and has rounder isophotes.

The disk is usually described by a 3D broken exponential profile:
\begin{equation}
    j(R,z)= j_0 \sech^{2/n}\left(\frac{nz}{2z_\mathrm{d}}\right) \times
\begin{cases}
  \exp\left(-\displaystyle\frac{R}{h_1}\right) , \; R \leq R_\mathrm{T}, \\
  \exp\left(\displaystyle-\frac{R}{h_2}\right)\exp\left(\displaystyle\frac{R_\mathrm{T}}{h_2}-\displaystyle
                       \frac{R_\mathrm{T}}{h_\mathrm{1}}\right),\; R > R_\mathrm{T},
\end{cases}
\label{eq:disk}
\end{equation}
where $j_0$ is the central luminosity density,
$z_0$ is the vertical scale height, and
$n$ is the parameter describing how strongly the disk is concentrated in the vertical direction.
The scale lengths $h_1$ and $h_2$ are the inner and outer radial scale lengths of the disk, respectively,
and $R_\mathrm{T}$ is the location of the disk break. In some cases, a third scale $h_3$ (a double-broken profile) is
introduced to describe an additional outer truncation of
the surface-brightness profile in the galaxy outskirts.
For the decompositions of edge-on galaxies with a B/PS bulge, the use of broken profiles
often becomes a necessity. In fact, the break in the profile frequently represents an inner truncation,
the existence of which is fully understandable from a dynamical point of view.
Since a bar naturally forms from the disk material,
a single exponential disk cannot extend all the way to the center;
therefore, the profile must be modified to adequately
represent real galactic disks. A similar model was used in~\cite{Brokendisk}
to describe the structure of CALIFA-DR3 (Calar Alto Legacy Integral Field
Area Data Release 3) galaxies. However, in that work,
    the model was strictly two-dimensional (infinitesimal disk),
    whereas here, the luminosity density defined by~(\ref{eq:disk})
    is integrated along the LoS to
    produce the proper disk image.
    By allowing the inclination to vary freely during fitting,
    we found that 75\% of our EGIPS
galaxies have an inclination greater than 85$^\circ$, while
    the remaining 25\% have inclinations
between 75$^\circ$ and 85$^\circ$.
Also note that, to compare the sizes of B/PS bulges with the
    underlying disk sizes, we use an equivalent of disk scale lengths,
defined as
$h_\mathrm{d,90}=R_\mathrm{90}/3.89$ and $z_{d,90}=z_{90}/2.3$ (see Appendix~\ref{ap:disk_mess}),
where $R_\mathrm{90}$ is the radius containing 90\% of the total disk intensity
    and $z_{90}$ is the disk
height containing 90\% of the disk intensity, respectively.

\begin{figure}[H]
    \includegraphics[width=0.7\linewidth]{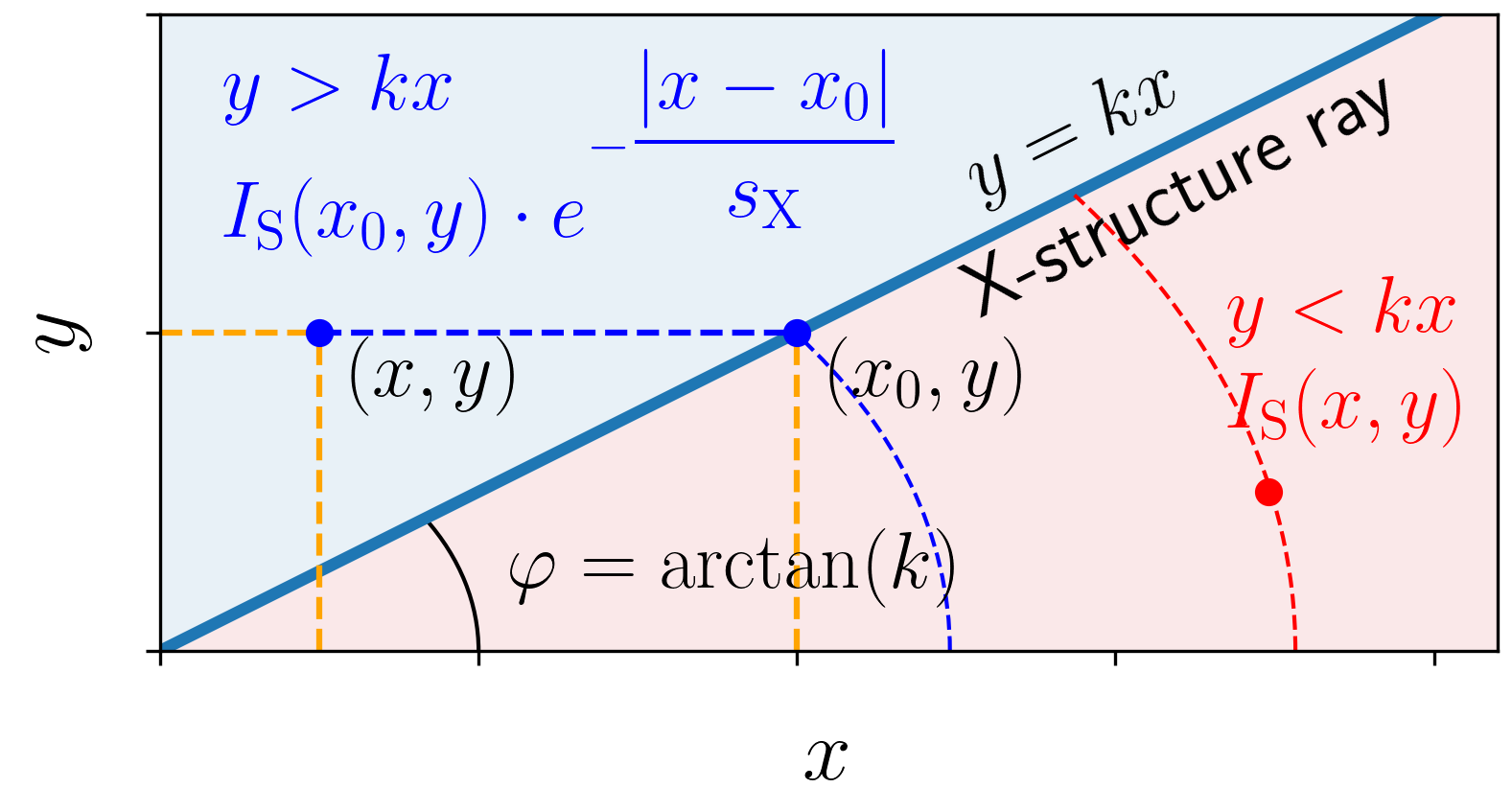}
    \caption{Schematic illustration of the intensity calculation
    in the X-shaped bulge photometric model
    used in the present study to represent the B/PS bulges of real
    and simulated galaxies. The figure is adapted from~\cite{Xold}
    with several modifications. The thick blue line indicates the X-structure ray;
    the light-red region marks the area below the ray,
        where the intensity is calculated following the standard
        S\'{e}rsic profile; the light-blue region marks the area above the ray,
        where the density truncation with the scale length $s_\X$ is introduced.}
    \label{fig:Xmodel}
\end{figure}

To properly represent the B/PS bulges, we employ an X-shaped bulge function
suggested in our previous work~\cite{Xold}. The model is a modified 2D S\'{e}rsic function
with an exponential truncation
added above the rays of X-structures:
\begin{equation}
I(x,y) =
\begin{cases}
I_\mathrm{S}(x,y), \, y \leq kx, \\
I_\mathrm{S}(x_0,y) \cdot \exp (-|x-x_0|/s_\mathrm{X}), \, \, y>kx.
\end{cases}
\label{eq:X}
\end{equation}
Here, $I_\mathrm{S}(x, y)$ is the intensity of S\'{e}rsic 2D function~(\ref{eq:ser}), $\varphi=\arctan (k)$
describes the slope of the X-structure rays relative to the disk plane, and the coordinate $x_0$ is the abscissa
of the point where the horizontal line $(y)$ intersects the X-structure ray in a given quadrant.
\textcolor{black}{A schematic illustration of how
the model intensity is calculated is presented in Figure~\ref{fig:Xmodel}.
We also note that the model has been slightly updated compared to
the version used in~\cite{Xold}: specifically, the intensity modifier
in the bottom row of Equation~(\ref{eq:X}) is now fixed at $I_\mathrm{S}(x_0,y)$
    instead of $I_\mathrm{S}(x,y)$. Additional details are
    provided in Appendix~\ref{ap:Xmodel}.}
    
The X-shaped bulge function was thoroughly discussed in~\cite{Xold}
and was tested on ``naked'' B/PS
bulges of numerical models, which can be obtained by subtracting the disk
from the galaxy using dynamical considerations (e.g., orbital
frequencies), full models with the disk, and a sample of real galaxies.
Essentially, this function introduces two additional parameters
compared to the S\'{e}rsic function:
the angle of the X-structure rays ($\varphi$)
    and scale length of the density decay from the ray along the
    horizontal direction towards the center ($s_\X$).
These parameters allow the model to describe a variety of morphologies,
including the characteristic peanut shape, i.e., the gap between the rays (if present).


The X-shaped model offers an advantage over models based on generalized ellipses,
which, while capable of modeling boxy
isophotes, cannot properly reproduce the gap between the rays.
However, we should note that our photometric function makes several simplifications,
such as assuming that the X-structure rays are straight
and relatively thin, and that the entire B/PS bulge has 4-fold symmetry
about the galaxy center.
In this study, as well as in our previous work, we also assume a S\'{e}rsic index of $n=1$.
This assumption simplifies the model and helps avoid potential degeneracies when other components,
such as the classical bulges, are present in the galaxy center.
While these assumptions are limitations of the model, trying to account for these complexities within
the framework of photometric decomposition would require additional parameters to describe the B/PS bulges.
In practice, a model with only two extra parameters is already challenging to fit,
as it requires an extensive search through the parameter space.
Therefore, this model keeps a balance between simplicity and an adequate description of the B/PS bulge component.

\begin{figure}[H]
    \includegraphics[width=0.9\linewidth]{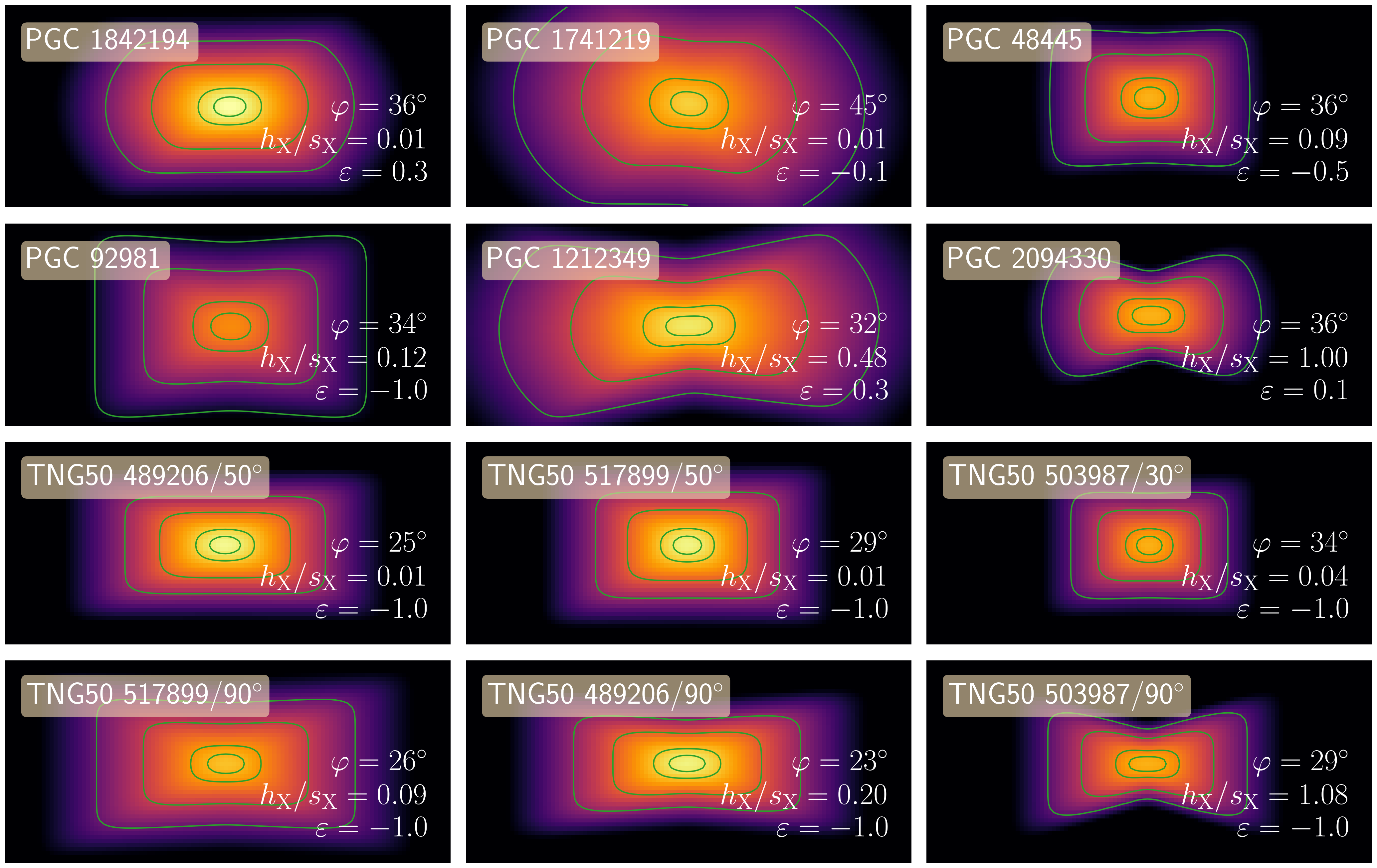}
    \caption{Examples 
 of the best-fit model of the X-shaped bulge
    with different values of $\varphi$, $h_\X/s_\X$, and $\varepsilon$
        for selected galaxies from EGIPS and TNG50.
        Green contours represent the isophotes corresponding to
        $0.02, \,0.1,\,0.5$, and $0.8$ of the intensity maximum.
        For TNG50, the text following the backslash
        indicates the bar's viewing angle.}
    \label{fig:X_ex}
\end{figure}

Some additional notes are as follows.
\textcolor{black}{In the course of this work, we characterize B/PS bulges
in terms of the ratio $h_\X/s_\X$, where by $h_\X$ we denote
the scale length of
the S\'{e}rsic part of Equation~(\ref{eq:X}), i.~e.,
the scale length of the profile below the X-structure ray. In practice,}
the ratio $h_\X/s_\X$ indicates the degree of
boxiness: small values correspond to horizontal isophotes above the ray (boxy),
while large values correspond to sharp peanut-shaped profiles.
The ellipticity is redefined as
\begin{eqnarray}
\epsilon =
\begin{cases}
1 - b/a, \, b \leq a \\
a/b - 1, \, b > a,
\end{cases}
\label{eq:ell}
\end{eqnarray}
where $a$ and $b$ are the major and minor axes of an elliptical isophote.
For $0\leq\epsilon<1$, this is the usual ellipticity,
while values $\epsilon<0$ correspond to elliptical
isophotes elongated perpendicular to the disk.
Specifically, $\epsilon=-1$ corresponds to vertical isophotes below the ray.
This flexibility in defining ellipticity allows
the model to properly reproduce boxy isophotes below the ray as well.
\textcolor{black}{Some examples of how the model looks depending on
    $\varphi$, $h_\X/s_\X$, and $\varepsilon$ are provided in Figure~\ref{fig:X_ex}.}

For galaxies with bright density enhancements in the disk plane and outside the B/PS bulge,
we also employ a 3D Gaussian ring component:
\begin{equation}
j(r,z) = j(r_\mathrm{ring},0) \exp\left(-\frac{(r - r_\mathrm{ring})^2}{2 \sigma_\mathrm{ring}^2}\right) \exp\left(-\frac{|z|}{h_z}\right),
\end{equation}
where $r_\mathrm{ring}$ is the ring radius, $\sigma_\mathrm{ring}$ is the ring thickness,
$h_z$ is the scale height, and $j(r_\mathrm{ring},0)$ is the luminosity density at $r=r_\mathrm{ring}$.


\subsection{Fitting Essentials}

For all galaxies in the sample and for the TNG50 cosmological models,
we obtained suitable multi-component photometric models using the pipeline described above.
In this subsection, we aim to provide the reader with an
overview of how this was achieved and what steps were taken to assess the quality
of the fits.

\begin{figure}[H]
    \includegraphics[width=0.95\linewidth]{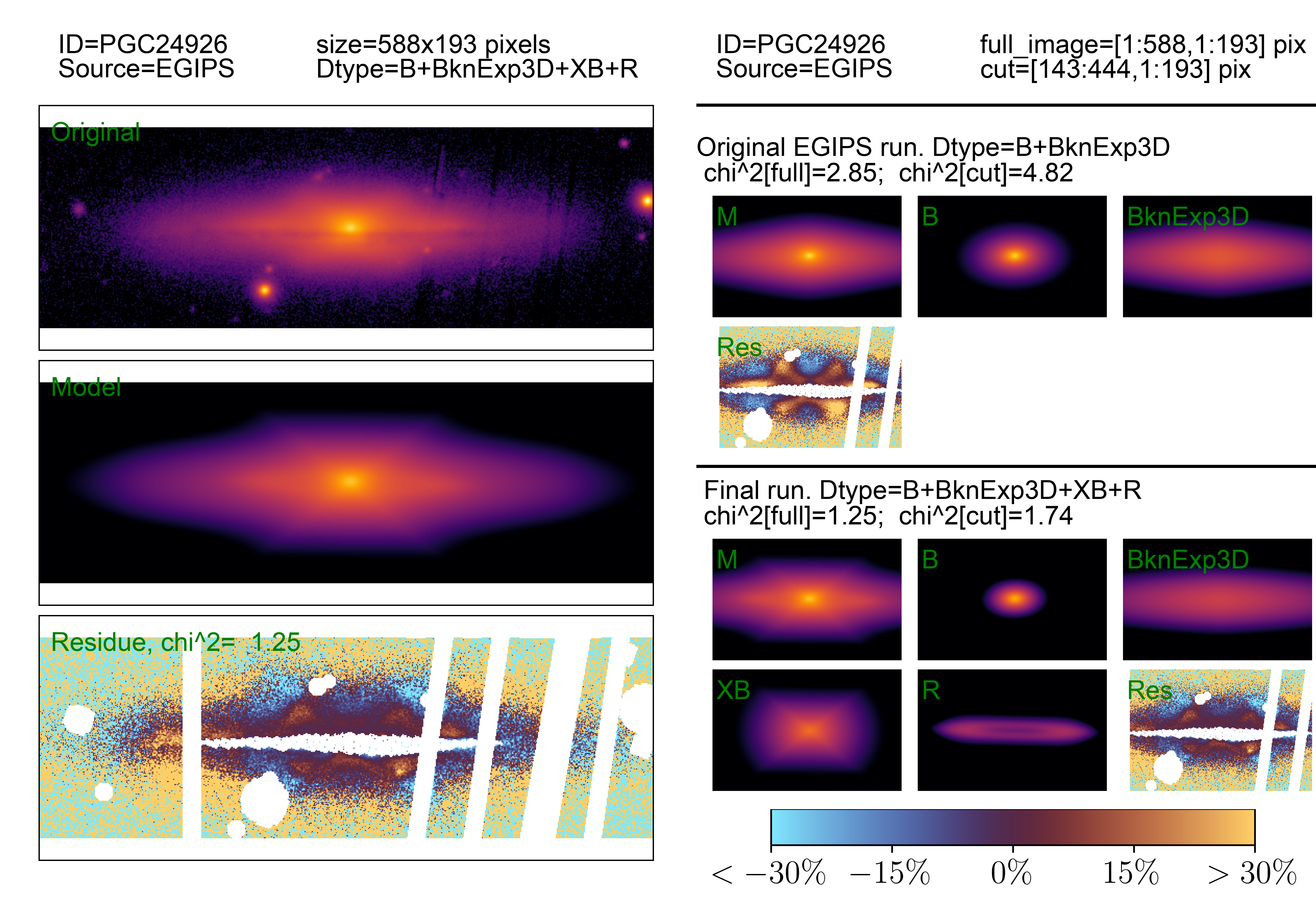}
    \caption{
The photometric decomposition of PGC~24926 (NGC~2690).
       (\textbf{Left}): 
       cropped and rotated Pan-STARRS
        $i$-band galaxy image
        (\textit{top}), 
 the best-fit photometric model (\textit{middle}), and the
        residual image (\textit{bottom}). (\textbf{Right}): the images of the
        individual photometric components and the residual image for the original
        EGIPS decomposition run with a simple S\'{e}rsic bulge (\textit{top})
        and decomposition performed in the present work
        with the X-shaped bulge model given by Equation~
        (\ref{eq:X}) and
        a Guassian ring (\textit{bottom}).
        The auxiliary text above the images provides various details of the decomposition.
        The images in the right
  column show a smaller area where the B/PS bulge resides. Two chi-square values
        are provided in auxiliary text: one is calculated over all pixels used for the decomposition
        (``chi\textasciicircum2[full]'') and the second one calculated over pixels
        from this smaller area (``chi\textasciicircum2[cut]'').
        The latter is more sensitive to the residuals in the central area of the galaxy.
        }
    \label{fig:dec_ex}
\end{figure}

An example of the final fit, relative residuals,
and images of the individual components for PGC~24926 (NGC~2690), a prominent galaxy with a B/PS bulge,
are presented in Figure~\ref{fig:dec_ex}.
To illustrate the improvement of the final model with the X-shaped bulge component
compared to the simple disk plus S\'{e}rsic bulge decomposition, the figure also
shows the corresponding photometric model from the original EGIPS run,
its residuals, and images of both the disk and the bulge components in the top right corner.
We prepared such comparative image collages for each galaxy in
our sample as part of our decomposition routine to assess the quality of the fit
(see the Supplementary Materials).
Each collage was checked to ensure that the X-shaped structure in the residuals,
if not entirely eliminated, became much less pronounced.

Figure~\ref{fig:dec_ex} illustrates specific aspects of our study and the data as well.
For example, note the vertical stripes that cross the image of the galaxy itself.
These are artifacts from the Pan-STARRS dataset and were generally masked manually.
The dust lane here is very prominent and rather thick,
and it was also masked during the decomposition.
The residual image from the bulge plus disk decomposition shows very prominent X-structures,
which become much less pronounced in the final model.
The residuals in the B/PS bulge region are reduced to less than 10\%.
There is also a central bulge with a much smaller scale length than the B/PS bulge
(3 arcsec vs. 9 arcsec, respectively)
and about the same S\'{e}rsic index of $n=1$.
Additionally, there are density enhancements corresponding to either a ring or bar spurs,
which appear in the residuals if no ring is included.
The inclusion of the ring significantly reduces the residuals, as expected.
Overall, as indicated in the legend of Figure~\ref{fig:dec_ex},
the model improves substantially, with the total
$\chi^2$ reduced by about three times,
both for the entire image and for the central cut,
where the B/PS bulge and the ring roughly reside.

\begin{figure}[H]
    \includegraphics[width=0.7\linewidth]{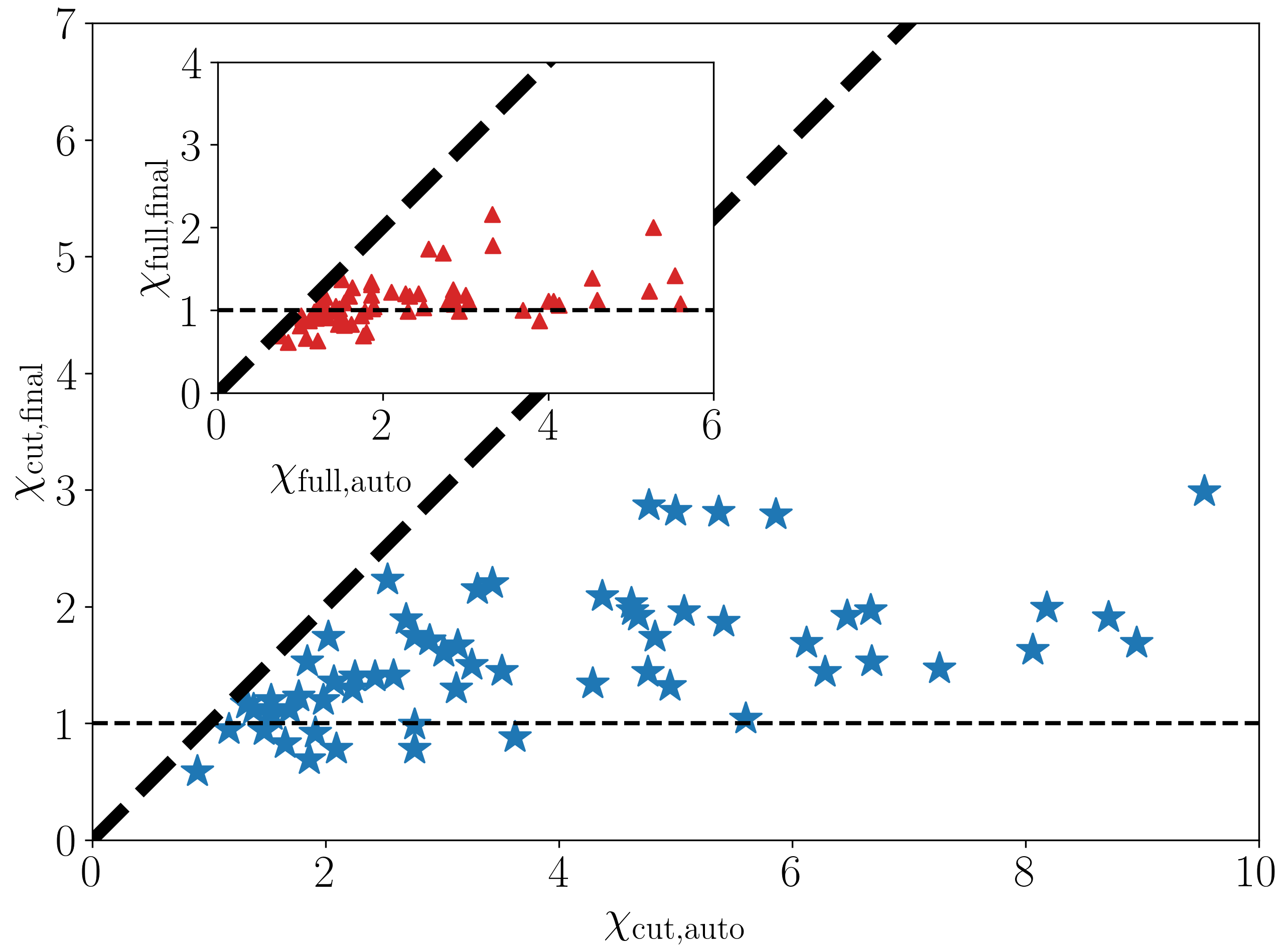}
    \caption{Comparison 
 of the $\chi^2$ statistic value from the original automatic EGIPS
    decomposition run with only bulge and disk components ($y$ axis) and the multicomponent
    decomposition performed in the present work ($x$ axis)
        for central area where the B/PS bulge resides (blue stars) and
        for the full galaxy image (inset, red triangles).}
    \label{fig:sigma}
\end{figure}

Figure~\ref{fig:sigma} compares the $\chi^2$ statistics
for all galaxies in our EGIPS sample between
the original two-component decomposition (disk plus bulge)
and our improved models for the central cut as well as the full galaxy image (see inset).
\textcolor{black}{The cut limits were determined visually and
correspond to the central region of the galaxy where the B/PS bulge roughly
resides (exact cut limits are provided in the corresponding
collages for each galaxy in the Supplementary Materials).
While the choice of limits is subjective,
    it provides a more sensitive metric
    for changes in the photometric model
    of the central component compared to the $\chi^2$
    calculated over the entire galaxy image.}
As can be seen from Figure~\ref{fig:sigma}, for all models, we indeed obtain a substantial improvement of the residue.
Basically, almost all models end up \mbox{with $1\lesssim \chi^2 \lesssim3$.}

\section{B/PS Bulge Parameters from Photometric Decomposition}
\label{sec:res}

In this section, we investigate the parameters of the
B/PS bulges obtained for both the real galaxies from EGIPS
and our TNG50 sample. We explore the possible correlations
between the B/PS parameters and other structural parameters of
galaxies, and also compare the parameters of real galaxies
with those of the TNG50 models.

We mainly focus on the following physical quantities:

\begin{enumerate}
    \item The X-structure opening angle $\varphi$.
    The angle value is generally linked to the dominant orbital family
    forming the B/PS bulge~\cite{Parul_etal2020}, which in turn is determined by the
    physical properties of the galaxy.
    However, it is also subject to projection effects,
    i.e., the orientation of the bar major axis with respect to
    the LoS~\cite{Smirnov_Sotnikova2018,Xold}.
\item The scale length of the S\'{e}rsic profile below the ray,
    which we denote as $h_\X$ to distinguish it
    from other scale lengths. Since we fixed S\'{e}rsic index to $n=1$ for our X-shaped bulge
photometric model, $h_\X$ corresponds a simple exponential scale length.
    (note that $r_e$ in Equation (\ref{eq:ser}) is always the effective radius, even for $n=1$.
    The corresponding exponential scale length $h_\X$ can be calculated as $r_e/1.678$.).
    Its value generally tells
    us how large the B/PS bulge is.
\item Other geometric properties of the B/PS bulge:
    shape of isophotes below the ray $\varepsilon$
    and the amplitude of the intensity dip between the rays,
    quantified by
    $h_\X/s_\X$ (see Section~\ref{sec:phot_func}).
    In general, $h_\X/s_\X$ may serve as an additional indicator of the bar viewing angle,
    with smaller values corresponding to a
    bar rotated closer to the LoS.
\end{enumerate}

Median values of the parameters for each subsample discussed here are
presented in Table~\ref{tab:params}
\textcolor{black}{with the exception of $h_\X/s_\X$ and $\varepsilon$,
which, as we show below, tend to degenerate $h_\X/s_\X \rightarrow0$
    and $\varepsilon\rightarrow-1$, respectively.}
We also estimate the typical spread of values in each subsample
as half of the interquartile range $(q_{75}-q_{25})/2.0$.

\begin{table}[H]
        \caption{Median values of the parameters for various photometric
    components for different subsamples
    considered in the present work. First column: the parameter under consideration,
    specifically, 
        $\varphi$---the opening angle of X-structure rays;
        $h_\X$---scale length of the B/PS bulge;
        $h_\X/h_\mathrm{d,90}$---B/PS bulge scale length relative to the disk scale length
        $h_\mathrm{d,90}$ (see Appendix~\ref{ap:disk_mess});
        $z_\mathrm{d,90}$---disk vertical scale length;
    $z_\mathrm{d,90}/h_\mathrm{d,90}$---disk thickness;
    $X/T$---B/PS bulge-to-total luminosity ratio;
    $X/D$---B/PS bulge-to-disk luminosity ratio;
    $B/T$---contribution of other bulges (excluding the B/PS bulge) to the total luminosity;
    $(B+X)/T$---total contribution of the central component (B/PS bulge plus other bulges).
    Columns from the second to fifth: median values and characteristic spread for
    EGIPS galaxies (second column) and TNG subsamples with different bar viewing angles
        (third to fifth columns, respectively).
    The spread of values (the values after the $\pm$ sign)
        is estimated as half of the interquartile range,
        $(q_{75}-q_{25})/2.0$.}
    \label{tab:params}
    \begin{tabularx}{\textwidth}{LCcCC}
    \toprule
    \textbf{Parameter} & \textbf{EGIPS}  & \textbf{TNG50 Bar} \boldmath{$90^\circ$} & \textbf{Bar} \boldmath{$50^\circ$} &
    \textbf{Bar} \boldmath{$30^\circ$} \\
    \midrule
    $\varphi$, deg & $37.13 \pm 3.11$ & $26.39 \pm 1.46$ & $30.64 \pm 3.21$ & $41.87 \pm 7.21$\\
$h_\mathrm{X}$, kpc & $1.10 \pm 0.31$ & $0.87 \pm 0.13$ & $0.74 \pm 0.14$ & $0.67 \pm 0.13$\\
$h_\mathrm{X}/h_\mathrm{d, 90}$ & $0.26 \pm 0.04$ & $0.26 \pm 0.07$ & $0.22 \pm 0.07$ & $0.17 \pm 0.04$\\
$h_\mathrm{d,90}$, kpc & $4.45 \pm 0.99$ & $3.73 \pm 0.68$ & $4.07 \pm 0.69$ & $4.07 \pm 0.79$\\
$z_\mathrm{d,90}$, kpc & $0.93 \pm 0.30$ & $0.99 \pm 0.21$ & $0.99 \pm 0.21$ & $1.05 \pm 0.23$\\
$z_\mathrm{d,90}/h_\mathrm{d, 90}$ 
& $0.23 \pm 0.04$ & $0.26 \pm 0.07$ & $0.25 \pm 0.10$ & $0.24 \pm 0.08$\\
$X/T$ & $0.22 \pm 0.04$ & $0.15 \pm 0.07$ & $0.17 \pm 0.07$ & $0.11 \pm 0.07$\\
$X/D$ & $0.36 \pm 0.10$ & $0.24 \pm 0.13$ & $0.32 \pm 0.16$ & $0.29 \pm 0.11$\\
$B/T$ & $0.08 \pm 0.09$ & $0.17 \pm 0.06$ & $0.16 \pm 0.06$ & $0.14 \pm 0.07$\\
$(B+X)/T$ & $0.34 \pm 0.08$ & $0.30 \pm 0.11$ & $0.31 \pm 0.12$ & $0.31 \pm 0.10$\\
\bottomrule
    \end{tabularx}

    \end{table}

\subsection{X-Structure Angles and B/PS Bulge Sizes}

\begin{figure}[H]
    \includegraphics[width=0.8\linewidth]{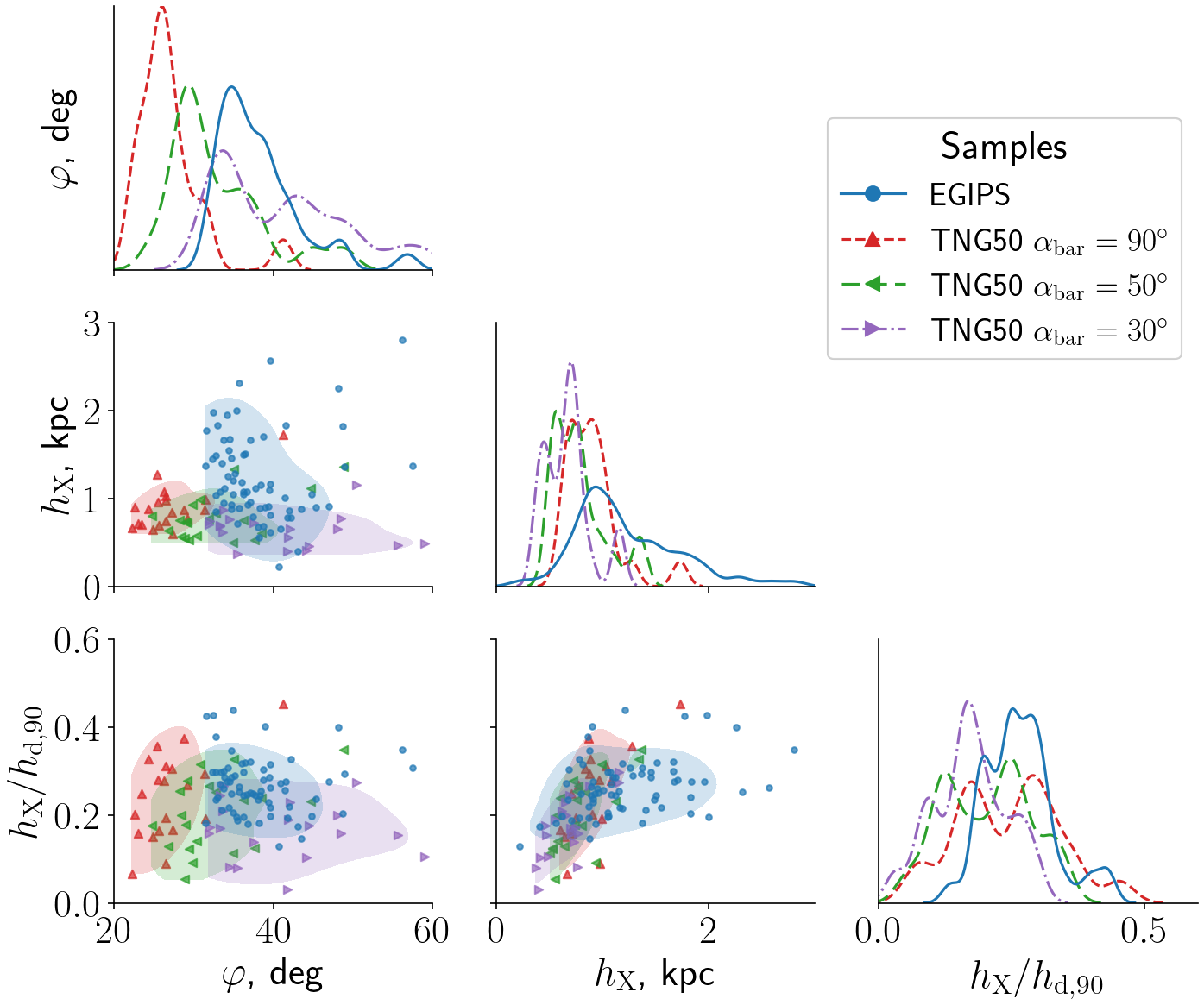}
    \caption{One-dimensional distributions of X-structure angles $\varphi$ and
    B/PS bulge scale lengths $h_\X$ (in kpc and in units of the
    disk scale length $h_\mathrm{d,90}$; diagonal panels),
        along with the corresponding correlation plots (off-diagonal panels),
        as measured from the photometric decomposition.
    Blue dots correspond to real galaxies from EGIPS sample,
    while red, green, and magenta triangles correspond to TNG50 subsamples with different
    bar viewing angles, 90$^\circ$, 50$^\circ$, and 30$^\circ$, respectively.
    Filled contours enclose 75\% of the respective subsamples.}
    \label{fig:phi_h}
\end{figure}

Figure~\ref{fig:phi_h} shows the distributions and correlation plots of measured
X-structure angles and the bulge scale length $h_\X$, both in absolute values (kpc)
and relative to the disk scale length $h_\mathrm{d, 90}$.
As can be seen from the figure, the angles $\varphi$ are mainly distributed between
$30^\circ$ and $50^\circ$ for EGIPS galaxies. This range is consistent with previous works,
both observational studies~\cite{Savchenko_etal2017, Xold} and numerical simulations of
isolated $N$-body galaxies~\cite{Smirnov_Sotnikova2018}.
The TNG50 galaxies are particularly interesting in terms of $\varphi$,
as we measure very small angles \mbox{$\varphi\sim22^\circ-23^\circ$} for some of them.
It is difficult to estimate how many X-structures with such small angles
are missed in observations due to selection effects,
as smaller-angle X-structures are generally harder to detect.
However, such small angles were not observed even in the isolated
galaxy simulations from~\cite{Smirnov_Sotnikova2018}, where the smallest angle was $25^\circ$,
found in the model with a substantial contribution from the non-baryonic component (dark halo)
and at late stages of evolution ($t > 6$ Gyr).

The linear size subplots in Figure~\ref{fig:phi_h} (\textit{middle} 
 rows and columns) reveal several important trends.
First, we observe that the higher the X-structure angle, the smaller the scale
length of the bulge in general.
One-dimensional distributions for different bar viewing angles in the TNG50 galaxies
show that this dependency is a natural consequence of projection effects: the closer
the bar is aligned with the LoS, the greater the X-structure angles is,
and the smaller the bulge appears to be (see also Table~\ref{tab:params}).

Second, the linear scales for real galaxies span a range from approximately 0.3 kpc to 2.0 kpc.
Comparing the scales of the TNG50 galaxies with our EGIPS sample, we see that the latter are generally
larger. The peak of the one-dimensional distribution of real galaxies is clearly shifted
toward greater values compare to TNG50 subsamples. In our TNG50 sample, there are almost
no B/PS bulges with scales greater than 1.5 kpc. However, such bulges are present in real galaxies.
We should note that the systematic absence of large bulges in our TNG50 sample due to selection effects is unlikely,
since larger bulges are easier to distinguish.
In terms of relative disk scales (\textit{third} row and \textit{third} column of Figure~\ref{fig:phi_h}),
the X-shaped bulges of real galaxies are also clearly larger,
with the peak of the distribution occurring at around $h_\X/h_{d,90}\sim0.26$.

\begin{figure}[H]
    \includegraphics[width=0.9\linewidth]{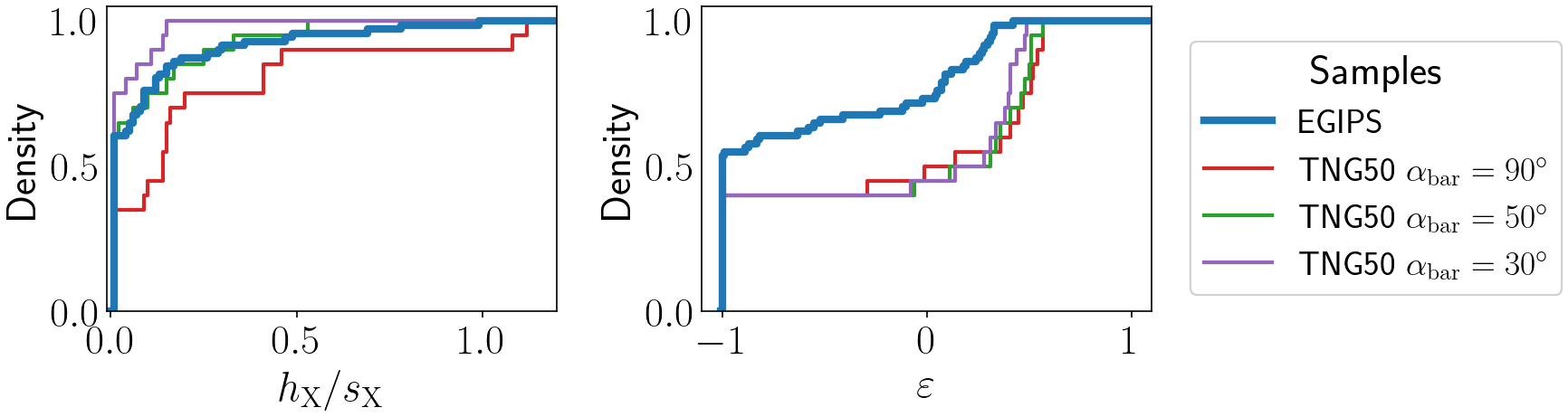}
    \caption{Cumulative distribution of isophotal parameters $h_\X/s_\X$ and $\varepsilon$ of the
    B/PS bulges represented by a photometric model of the X-shaped bulge in EGIPS and
    TNG50 individual subsamples with different bar viewing angles.
    The color scheme follows that of Figure~\ref{fig:phi_h}.}
    \label{fig:Xiso}
\end{figure}

Figure~\ref{fig:Xiso} presents the cumulative distributions of isophotal
parameters for X-shaped bulges: $h_\X/s_\X$ (gap above the ray) and $\varepsilon$ (isophotal profile below the ray).
For the parameter $h_\X/s_\X$, we find that for most real galaxies and TNG50 models,
this parameter tends to degenerate to $h_\X/s_\X\rightarrow0$. This indicates that, for most of the studied galaxies and models,
the isophotes above the rays are horizontal (boxy profile). For isophotes below the ray,
we observe an interesting
distinction between the bulges in TNG50 and EGIPS galaxies.
In TNG50 galaxies, the isophotes are more flattened ($\varepsilon>0$), while in EGIPS galaxies,
the X-shaped bulges are characterized
by more vertical isophotes ($\varepsilon<0$).

\subsection{B/PS Bulge and Disk}

\begin{figure}[H]
    \includegraphics[width=0.95\linewidth]{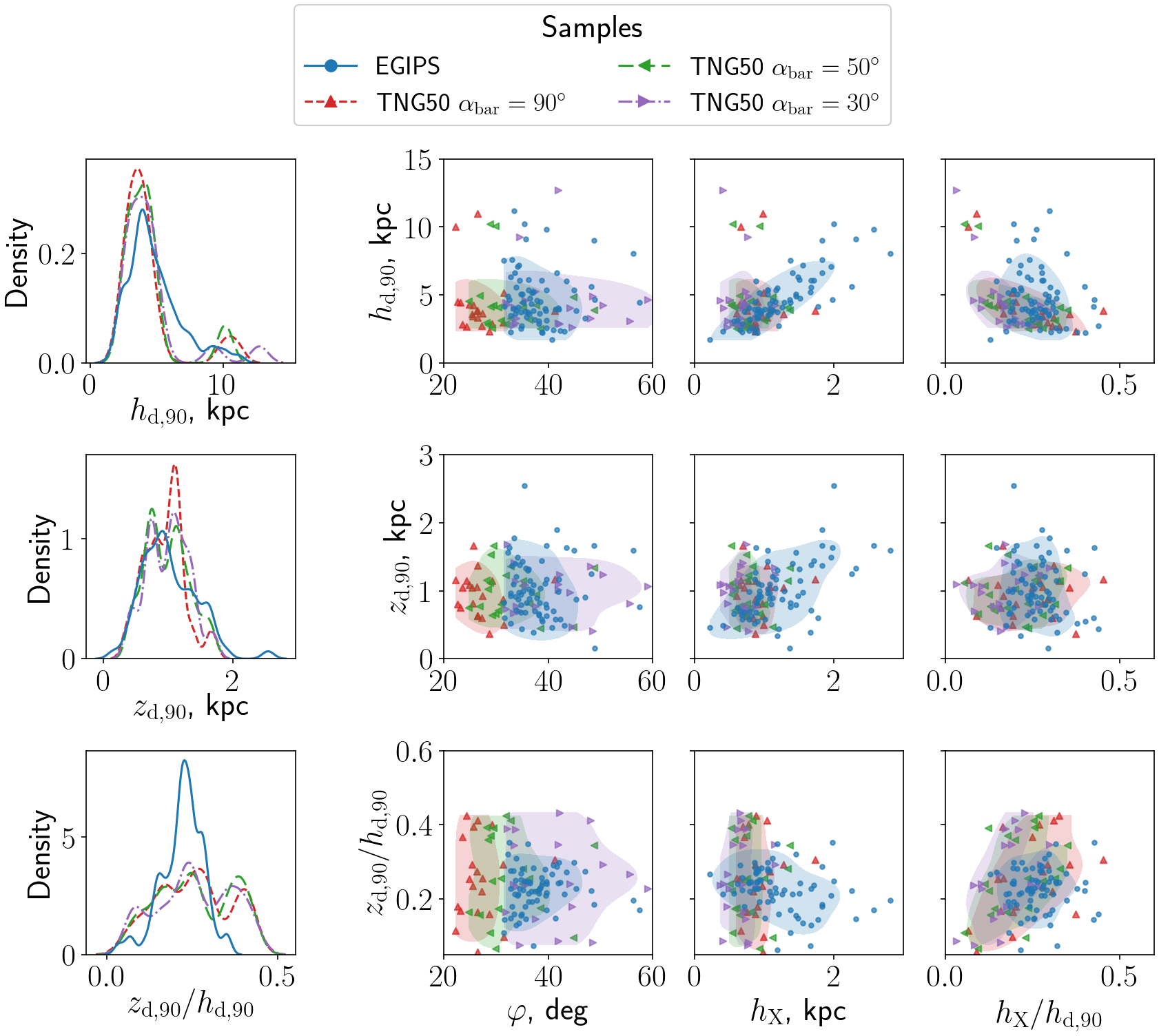}
    \caption{(\textbf{Left}): 
    1D distributions of disk scale lengths,
        in-plane (\textit{top}) 
 and vertical (\textit{middle}),
        and disk thickness (\textit{bottom}).
        (\textbf{Second to fourth columns}): disk scales (\textit{top} and
        \textit{middle} rows) and thickness (\textit{bottom} row),
        depending on the X-structure angles (\textbf{second} column) and B/PS bulge
        scale lengths (in kpc and relative to the disk scale length,
        \textbf{third} and \textbf{fourth} columns, respectively).
        The color and symbol notation follows that of Figure~\ref{fig:phi_h}.}
    \label{fig:Xdisk}
\end{figure}

Figure~\ref{fig:Xdisk} presents the distributions of the measured disk scales,
$h_\mathrm{d,90}$ and $z_\mathrm{d,90}$, as well as the disk thickness,
and the dependence of the B/PS bulge parameters on these quantities.
As the 1D distributions show (\textit{left} column), the disk scale length of the EGIPS galaxies
lies between 3 and 6~kpc for most of them.
For the TNG50 galaxies, the scale lengths
fall within the same range but are, on average, slightly smaller.
Also note the slight shift of the scale length to larger
values as the bar orientation moves closer to the LoS (see Table~\ref{tab:params}).
\textls[-15]{For the disk thickness, the distribution peaks at $h_\mathrm{d,90}/z_\mathrm{d,90}\sim0.23$,
which is roughly consistent with the value measured by~\cite{Mosenkov_2015}.
The disks of the measured TNG50 galaxies are generally slightly thicker,
as they tend to have smaller $h_\mathrm{d,90}$ while exhibiting approximately the \mbox{same $z_\mathrm{d,90}$}.}

Regarding the B/PS bulge parameters, we find that the angles do not show
a clear dependence on either the linear scales or the disk thickness.
At the same time, larger bulges are generally found in larger disks.
It should be noted that in our previous numerical study of a variety of models~\cite{Smirnov_Sotnikova2018},
we found that the angle should depend on the disk thickness;
however, we do not observe such a dependence here.
It is difficult to pinpoint a clear reason for that;
however, it should be noted that \cite{Smirnov_Sotnikova2018} compared the parameters
of X-structures depending on the original parameters of the disk (i.e., at the start of simulations).
It is quite possible that in observations we do not measure the same thickness, but
rather the thickness of an already evolved disk,
and therefore the dependence may no longer be apparent.

\subsection{B/PS Bulge Intensity Contribution}

\begin{figure}[H]
    \includegraphics[width=0.95\linewidth]{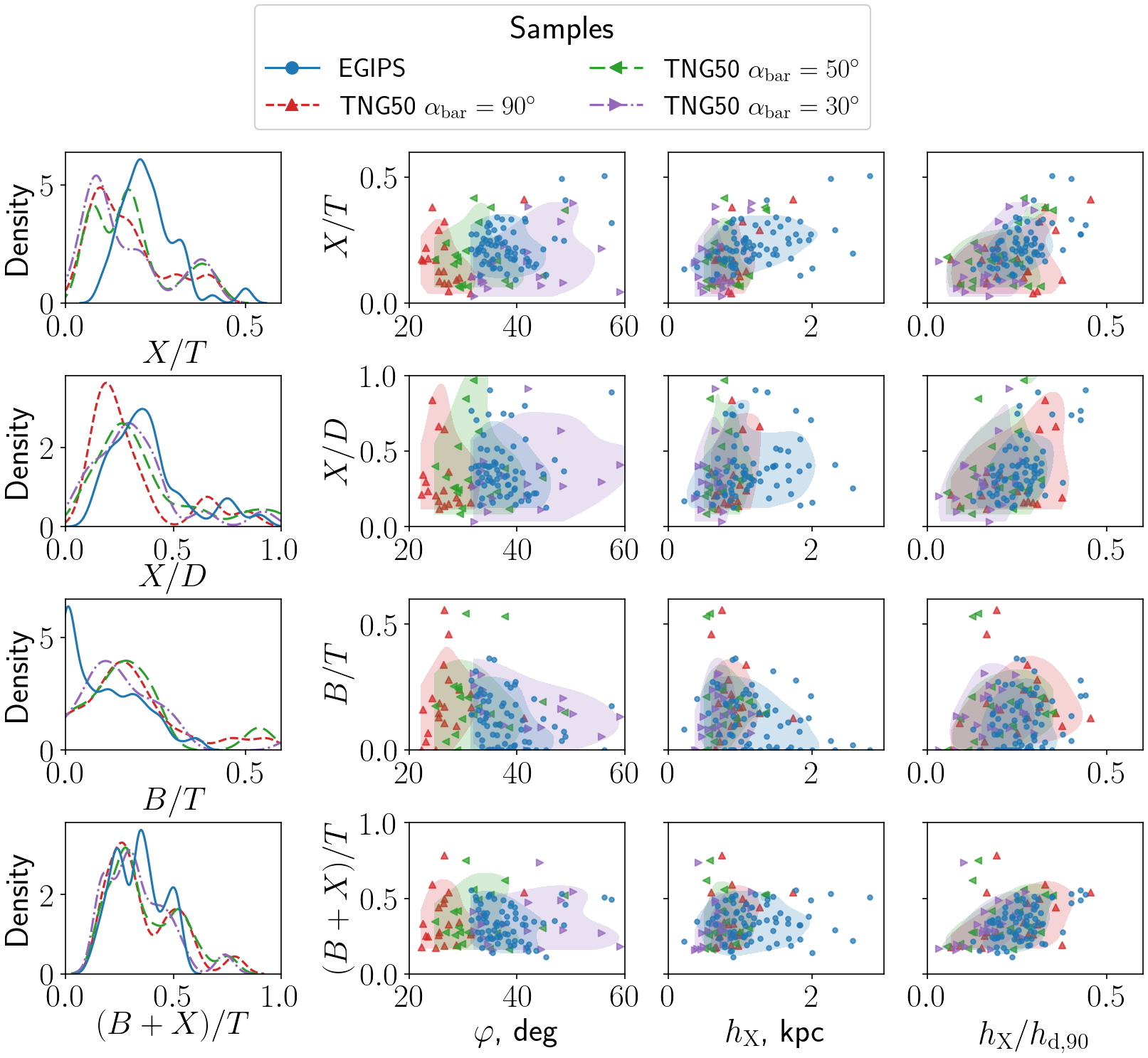}
    \caption{(\textbf{Left}): 
    1D distributions of B/PS bulge contribution to
    total intensity $X/T$ (\textit{top} row), 
    relative to the disk
    only ($X/D$, \textit{second} row), the contribution of other bulges
        $B/T$ (\textit{third} row), and the cumulative contribution of
        B/PS bulge and other bulges $(X+B)/T$ (\textit{bottom} row).
        (\textbf{Second to fourth columns}): the parameters shown in
        the left column of each row, depending on the X-structure angles
        (\textbf{second} column) and B/PS bulge scale lengths
        (in kpc and relative to the disk scale length,
        \textbf{third} and \textbf{fourth} columns, respectively).
        The color and symbol notation follows that of Figure~\ref{fig:phi_h}.
}
    \label{fig:Xint}
\end{figure}

Figure~\ref{fig:Xint} presents the distributions of the B/PS bulge contribution
to total intensity, denoted here as $X/T$ (\textit{top} row),
relative to the disk only ($X/D$, \textit{second} row),
the contribution of other bulges $B/T$, calculated as a sum of all other S\'{e}risc
photometric components (e.g., $(B_1+B_2)/T$ in the case of two additional bulges, \textit{third} row),
and the cumulative contribution of the B/PS bulge and other bulges $(X+B)/T$ (\textit{bottom} row).

For $X/D$ and $X/T$, we find that the B/PS bulges
of real galaxies are generally more prominent compared
to TNG50 galaxies (see also the values listed in Table~\ref{tab:params}).
For $X/D$ in EGIPS galaxies, the median value is 0.35,
which is consistent with the $B/D$ values measured in~\cite{EGIS} ($B/D\sim0.3$),
but appears to be larger than those measured in~\cite{Savchenko_etal2017,Yoshino_Yamauchi2015}, where the median $B/D$ values were
smaller ($B/D\lesssim0.2$). For $(B+X)/T$, we find that these cumulative values
can be quite large (up to 0.5), with the median value $\mu_{(B+X)/T}$
around 0.3 for all subsamples.

Note that the contribution of other bulges (denoted $B/T$ here) is, conversely,
greater in TNG50 galaxies. For instance,
comparing the median $B/T$ value in EGIPS and TNG50 galaxies,
where the bar is rotated perpendicular to LoS, $\mu(B/T)_{\mathrm{EGIPS}}=0.07$
and $\mu(B/T)_{\mathrm{TNG50}}=0.17$.
This is qualitatively consistent with our observations
during the decomposition of TNG50 galaxies.
If the inner bulges are not adequately represented in their photometric model,
the X-shaped bulge photometric model tends to degenerate into a standard 2D S\'{e}rsic profile.
The $B/T$ values of TNG50 galaxies are in good
agreement with the values measured in~\cite{Bulges_TNG},
where the median measured value was $\mu (B/T)=0.13$,
based on 1D exponential plus S\'{e}rsic fits.


As for the dependencies of the B/PS bulge parameters on the disk,
the only clear dependence is that $X/T$ correlates with
the size of the bulge (\textit{top} row of Figure~\ref{fig:Xint}).
Interestingly, we do not observe any dependence of the angles of X-structures on $B/T$ values.
In~\cite{Smirnov_Sotnikova2018}, we found that X-structures in galaxies with larger
$B/T$, assuming
$B$ corresponds to the classical bulge,
tend to be less flattened. However, Figure~\ref{fig:Xint} does not support \mbox{this conclusion}.

\section{Discussion}
\label{sec:discussion}

\subsection{X-Structure Angles: Physics or Projection Effects?}
One of the crucial questions in observational studies of B/PS bulges in general, and, naturally, in the present work, is what the observed angles of X-structures
can tell us about the galaxies in which these angles are measured.
On the one hand, simulation studies~\cite{Smirnov_Sotnikova2018, Parul_etal2020}
show that the angles of X-structures are tightly connected with the orbital
families that form the observed B/PS bulge. Because of this connection,
the X-structure angles should depend on the overall
gravitational potential of the system as well as on the bar properties.
From~\cite{Smirnov_Sotnikova2018}, we know that the X-structure angles
observed in side-on bars (bar viewing angle of 90$^\circ$) vary from at
least 25$^\circ$ to 40$^\circ$, depending on physical factors
such as bar age, the presence of a light or massive dark halo,
and whether the bar forms from an initially thin or thick disk.
On the other hand, the bar alignment with respect to the LoS
strongly influences the distribution of the observed angles.
As can be seen from Figure~\ref{fig:phi_h} and Table~\ref{tab:params} for the TNG models,
as well as from Figure~22 
in~\cite{Smirnov_Sotnikova2018},
the typical shift in the X-structure angle between a bar viewed side-on and
the same bar rotated to 30$^\circ$ is about 15$^\circ$.
This value is comparable to the intrinsic spread of
angles found for strictly side-on bars across different $N$-body simulations.

Table~\ref{tab:phi} presents the Spearman rank correlation coefficients
for the X-structure angles and various parameters and compares
the correlations with our qualitative expectation from numerical study
of~\cite{Smirnov_Sotnikova2018} for such correlations.
Specifically, we consider here the B/PS bulge scale lengths, both absolute
and relative to the disk scale,
the disk thickness, and B/PS bulge contribution both for EGIPS and
TNG50 samples. We also investigate separately the correlation between
the X-structure angles and the contribution of bulges
with S\'{e}rscic index $n>2$ for EGIPS, since~\cite{Smirnov_Sotnikova2018}
considered the impact of the classical bulges on the X-structures,
rather than the pseudobulges.
To facilitate comparison
with observation data,
we merged the results for TNG50 subsamples with different bar viewing angles here.
We also separately considered a subsample of EGIPS galaxies,
denoted as ``EGIPS (cut)'',
with measured disk inclinations $i>80^\circ$
and moderate X-structure angles $\varphi<45^\circ$.
Basically, the subsample formed this way should be more robust
to possible errors, associated with resolution and disk inclination.
The cut consists of 60 galaxies.

\begin{table}[H]
        \caption{Spearman 
 rank correlation coefficients between measured
    X-structure angles and various parameters of the B/PS bulge,
        disk, and other bulges (first column) for EGIPS galaxies (second column),
        a cut of the EGIPS sample (third column; see the main text),
        and TNG50 models (fourth column). The statistically significant correlations
            (\emph{p}-value < 0.05) are indicated in bold. 
 In the fifth column,
        we present our qualitative expectations based on the
        results of previous studies on X-structures and B/PS bulges.}
    \label{tab:phi}
    \begin{adjustwidth}{-\extralength}{0cm}
    \begin{tabularx}{\fulllength}{lcccC}
    \toprule
        \textbf{Parameter} & \textbf{EGIPS} & \textbf{EGIPS (Cut)} & \textbf{TNG50} &
    \textbf{Qualitive Expectations from Smirnov  \& Sotnikova
        (2018,~\cite{Smirnov_Sotnikova2018})} \\
    \midrule
    $h_\mathrm{X}/h_\mathrm{d, 90}$ & $-0.21$  & $\mathbf{-0.29}$  & $-0.06$ & \parbox[c][1.5cm][c]{4cm}
    {Strongly negative due to
    \\ 1.\,bar
    viewing
    angles\\2.\,secular bar growth}
    \\ \midrule
$h_\mathrm{X}$, kpc & $\mathbf{-0.30}$  & $\mathbf{-0.48}$  & $-0.17$ & 
                           Same as for $h_\X/h_\mathrm{d,90}$ but should additionally reflect
                           the systematic trend of  larger bars residing in more massive (larger)
                           galaxies, see~\cite{Erwin_Barsizes}
                           \\ \midrule
$z_\mathrm{d,90}/h_\mathrm{d, 90}$ & $0.07$  & $0.17$  & $0.02$ & Positive, see Figure~16 
 in~\cite{Smirnov_Sotnikova2018}\\ \midrule
$X/D$ & $-0.10$  & $\mathbf{-0.27}$  & $0.21$ &
                  Negative due to the bar growth, ideally should be independent from the bar viewing angle
                  \\ \midrule
$B{(n>2)}/T$ & $-0.17$  & $-0.10$  & -- & Positive, see Figure~16 in~\cite{Smirnov_Sotnikova2018} and Figure~9 in~\cite{Xold}\\
\bottomrule
    \end{tabularx}
\end{adjustwidth}
    \end{table}

In general, Table~\ref{tab:phi} highlights that there are no strong correlations
or anti-correlations between X-structure angles and
any of the considered parameters, except for the absolute B/PS bulge scale length $h_\X$,
even when we examine a
subsample of galaxies selected by disk inclination and moderate X-structure angles.
Let us discuss the correlations we can expect to observe for each of the considered parameters,
based on our current understanding of B/PS bulges.

For $h_\X/h_\mathrm{d,90}$, one would expect a strong negative correlation due
to projection effects (i.e., the B/PS bulge should appear smaller as the bar
aligns closer to the LoS). Additionally, in numerical simulations,
bars tend to slow down, grow in size over time,
and their X-structures tend to flatten.
The issue here is that no strong negative correlation is observed in the EGIPS sample.
There is also no strong negative correlation in the TNG50 data.
Figure~\ref{fig:phi_h} (left bottom subplot) gives a clue
as to why this correlation may be absent.
The figure shows that the intrinsic spread in B/PS bulge sizes plays a significant role,
when estimating the correlations.
For our TNG50 galaxies, the spread is large enough that it becomes
comparable to the shift in median values due to bar viewing angles.
For example, compare the intrinsic spread of values
for bars observed perpendicular to the LoS in TNG50
$\Delta (h_\X/h_\mathrm{d,90})=0.07$
with the shift in the median value between TNG subsamples (see Table~\ref{tab:params}),
$\mu(h_\X/h_\mathrm{d,90})_\mathrm{TNG\, 90^\circ} = 0.26$ versus
$\mu(h_\X/h_\mathrm{d,90})_\mathrm{TNG\, 30^\circ} = 0.17$.

\begin{figure}[H]
    \includegraphics[width=0.9\linewidth]{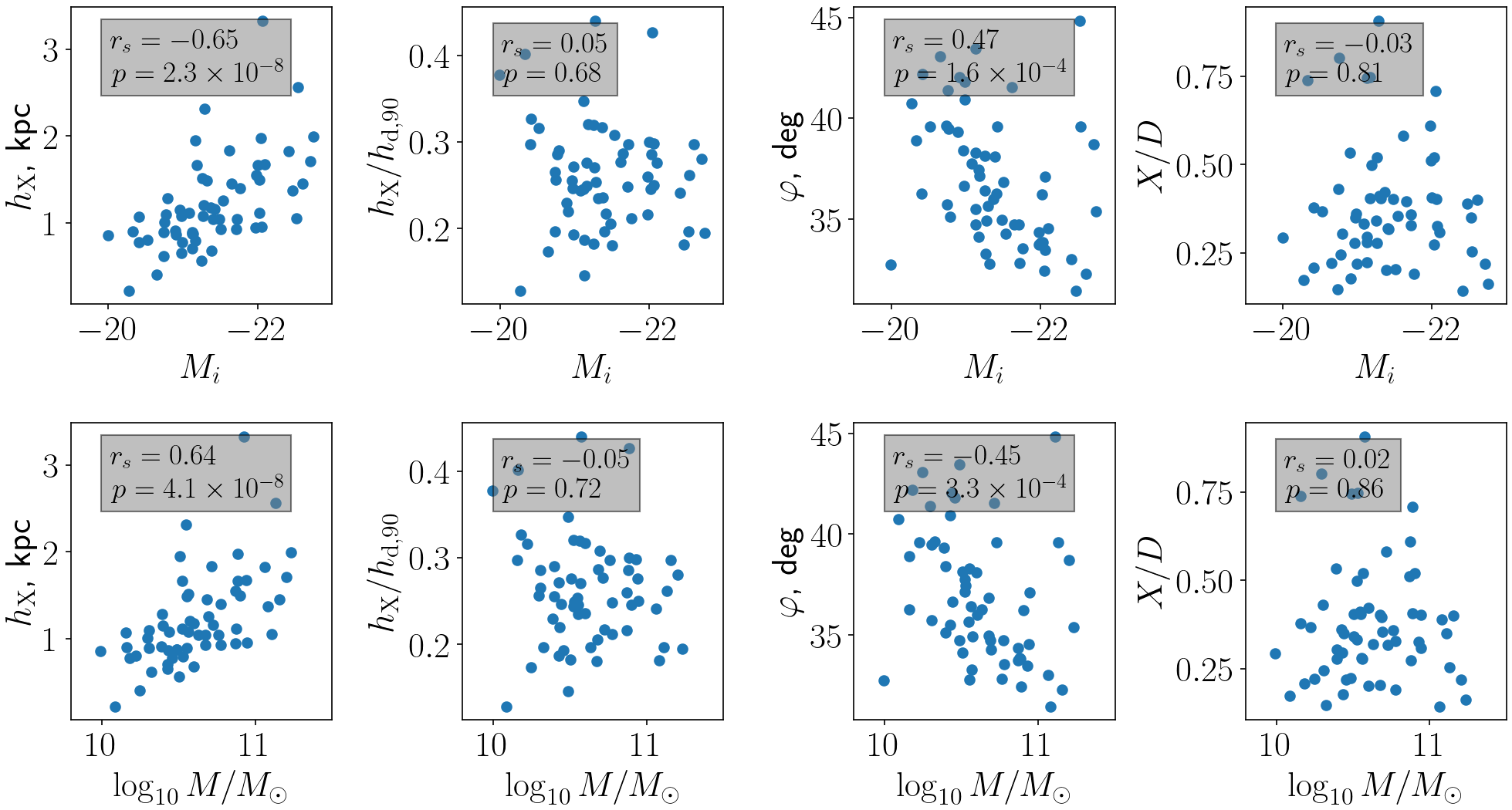}
    \caption{B/PS 
 bulge scale length in kpc (\textbf{first columns} 
    column),
        in units of the disk scale (\textbf{second} column),
        X-structure angles (\textbf{third} column), and B/PS bulge intensity
        contribution (\textbf{forth} column) depending on the absolute magnitude $M_i$ (\textbf{top} row)
        and galaxy mass (\textbf{bottom} row), respectively.
        The Spearman correlation coefficient and the corresponding \emph{p}-value 
        are provided in a textbox
        in each \mbox{corresponding subplot}.
}
    \label{fig:ring}
\end{figure}

For absolute value of B/PS bulge sizes $h_\X$ we see a strong
anti-correlation with X-structure angles.
Figure~\ref{fig:ring} explains why the anti-correlation for absolute
values of sizes can be stronger compared to relative sizes.
Here, we plot the sizes of B/PS bulge, X-structure angles,
and B/PS bulge intensity contribution depending
on the absolute magnitude $M_i$ of the galaxy in $i$-band
(\textit{top} row) and galaxy mass (\textit{bottom} row)
for our cut of the EGIPS sample.
The masses of the galaxies were roughly estimated based on calibrations
presented in~\cite{Mass_Calib}. Generally speaking,
both $M_i$ and the galaxy mass do not depend on the bar alignment to the
LoS, and, thus, free of projection effects.
In each individual
subplot, we also show a value of the Spearman correlation coefficient $r_s$ between
the corresponding parameters and absolute magnitude $M_i$ and the galaxy mass,
respectively.

Figure~\ref{fig:ring} reveals several trends.
First, the absolute sizes of B/PS bulges (\textit{left} column)
increase with galaxy mass, showing a significant positive correlation
($r_s \gtrsim 0.6$). This trend is consistent with two established facts:
the number of B/PS bulges in observations sharply rises for galaxies with
$M \gtrsim 10^{10.4-10.5}$~\cite{Erwin_Debattista2017, Marchuk2022},
and the sizes of bars, from which B/PS bulges originate,
also increase for galaxies with $M \gtrsim 10^{10.1}$~\cite{Erwin_Barsizes},
being larger in more massive galaxies. The left column of Figure~\ref{fig:ring}
shows that for galaxies with $M < 10^{10.4}$, B/PS bulges tend to be smaller,
which may explain why they are more likely to be missed in B/PS bulge samples
due to selection effects.
On the other hand, larger bars should host larger B/PS bulges.
Since larger bars are typically found in more massive galaxies,
this would naturally imply that B/PS bulges in these galaxies should also be larger,
which is supported by our measurements.
However, neither the relative sizes of B/PS bulges (\textit{second} column of
Figure~\ref{fig:ring}) nor their relative contribution to
total intensity (\textit{fourth} column of Figure~\ref{fig:ring})
increase with galaxy mass. This absence of correlation suggests that,
while B/PS bulges are larger in more massive galaxies, they are not necessarily
older than those in less massive galaxies. If they were older,
we would expect the relative sizes and intensity contributions to increase as well,
which is not observed. Another trend,
which has not been highlighted in previous B/PS bulge studies,
is the moderate anti-correlation between X-structure angles and galaxy mass.
As shown in the third column of Figure~\ref{fig:ring}, X-structure angles tend to
be smaller in more massive galaxies. While this result requires further testing
with larger galaxy samples, it could suggest an intriguing distinction
in the orbital dynamics of larger versus shorter bars.

For the other parameters, specifically the disk thickness and the bulge-to-total ratio,
we observe a very weak positive correlation, while for the bulge-to-total ratio
the correlation is weakly negative.
However, both correlations have \emph{p}-values above 0.05
and are therefore not statistically significant. 
We also emphasize that there are no bulges
with $n>2$ in our TNG50 sample (see also below); therefore, the corresponding correlation
value is not included in the table. Regarding disk thickness, we note
that the largest value examined in~\cite{Smirnov_Sotnikova2018} was
$z_\mathrm{d}/h_\mathrm{d}=0.2$, whereas both EGIPS and TNG50 disks are generally thicker.
Moreover, the model with $z_\mathrm{d}/h_\mathrm{d}=0.2$ from~\cite{Smirnov_Sotnikova2018}
exhibited a prolonged
buckling phase,
while the B/PS bulges studied here do not show any strong asymmetry.
This is likely due to the presence of additional bulge components,
which tend to stabilise the bar against buckling~\cite{NoBuck}.
In short, direct comparison between~\cite{Smirnov_Sotnikova2018} and either
the EGIPS or TNG50 data is difficult in terms of disk thickness,
as no models with $z_\mathrm{d}/h_\mathrm{d}\gtrsim0.2$ were explored in that study.
We conclude that further numerical studies, including models with thicker disks
and with contributions from additional bulge components, are required for a proper
investigations of trends in X-structure angles with disk thickness and other \mbox{bulges
contribution}.

\subsection{B/PS Bulges in TNG50}
Here, we want to share some additional thoughts regarding
B/PS bulges in Illustris TNG50 in light of the 2D photometric decomposition
carried out in the present work.

First, as we showed in Section~\ref{sec:res},
the B/PS bulges in TNG50 galaxies are smaller than
those in real galaxies, both in absolute terms and relative to the disk scale.
These findings can be viewed from the perspective
of the bar studies of TNG50 galaxies carried out so far. 
For example, ref. \cite{Frankel22} found that bars in MaNGA~\cite{Manga}
are larger than bars TNG50, and the corresponding sizes
differ by about 50\% in terms of the mean values.
At the same time~\cite{Habibi2024}, showed that bars in TNG50 do not display
the trend with galaxy mass that is observed in real galaxies.
Since B/PS bulges originate from the corresponding bars,
it is not surprising that our decomposition reveals
that B/PS bulges are also smaller in TNG50.
On the other hand, B/PS bulges in
TNG50 galaxies show rather small X-structure angles when
their bars are seen side-on.
The problem is that, based on orbital studies of B/PS simulations carried out
so far~\cite{Parul_etal2020}, we do not yet understand
which properties of the overall gravitational potential
could lead to the formation of X-structures with such small angles.
Further orbital studies of bars in TNG50 may help to answer this question.


The second point concerns the other bulges, except B/PS bulges,
in TNG50 and the overall decomposition
of TNG50 galaxies. During this study, we found that
TNG50 galaxies often have relatively bright central component(s)
located within the central region of the B/PS bulge.
Unless properly accounted for during the fitting,
these central components make it impossible to fit the B/PS bulge
model. This issue is reflected in Figure~\ref{fig:Xint} and Table~\ref{tab:params},
where the contribution of other bulges, denoted as $B/T$,
is found to be larger in TNG50 (see also the Supplementary Materials).
This distinction may be related
to the fact that we neglect the dust component when preparing
the TNG50 galaxies images with {SKIRT}.
Since the dust typically concentrates towards the galaxy center, it
would likely decrease the intensity of the central
component more than the disk contribution, leading to relatively smaller $B/T$.
On the other hand, some studies~\cite{Bulges_TNG} report that
bulges with high S\'{e}rsic indexes ($n>2$) are relatively rare in TNG50 galaxies,
which is inconsistent with observational data. In the current study,
we found that many central components of real galaxies exhibit S\'{e}rsic indexes $n>2$
(see below), whereas the central
components of TNG50 galaxies typically have $n\sim1$ and even smaller.
Since this study is purely photometric, identifying the reasons for these distinctions
is beyond the scope of the present work. However, understanding them
is crucial if we want to properly compare the structure of simulated and real galaxies.

\subsection{B/PS Bulge as a Part of Composite Bulges}

In most of our EGIPS galaxies, the central region is represented
by a S\'{e}rsic component together with the X-shaped bulge component.
In this sense, almost all of our bulges are
composite, similar to the case of NGC~4565 described by \cite{2B_Kormendy},
where a small inner bulge is found within the central part of a larger B/PS bulge.
An important question regarding such composite bulges,
emphasized by \cite{2B_Kormendy}, is how the bulge properties inferred
from a simple two-component (bulge + disk) decomposition compare
with the properties of the central bulge obtained from a more
sophisticated decomposition that explicitly
includes the B/PS bulge as a separate component.

\begin{figure}[H]
    \includegraphics[width=0.98\linewidth]{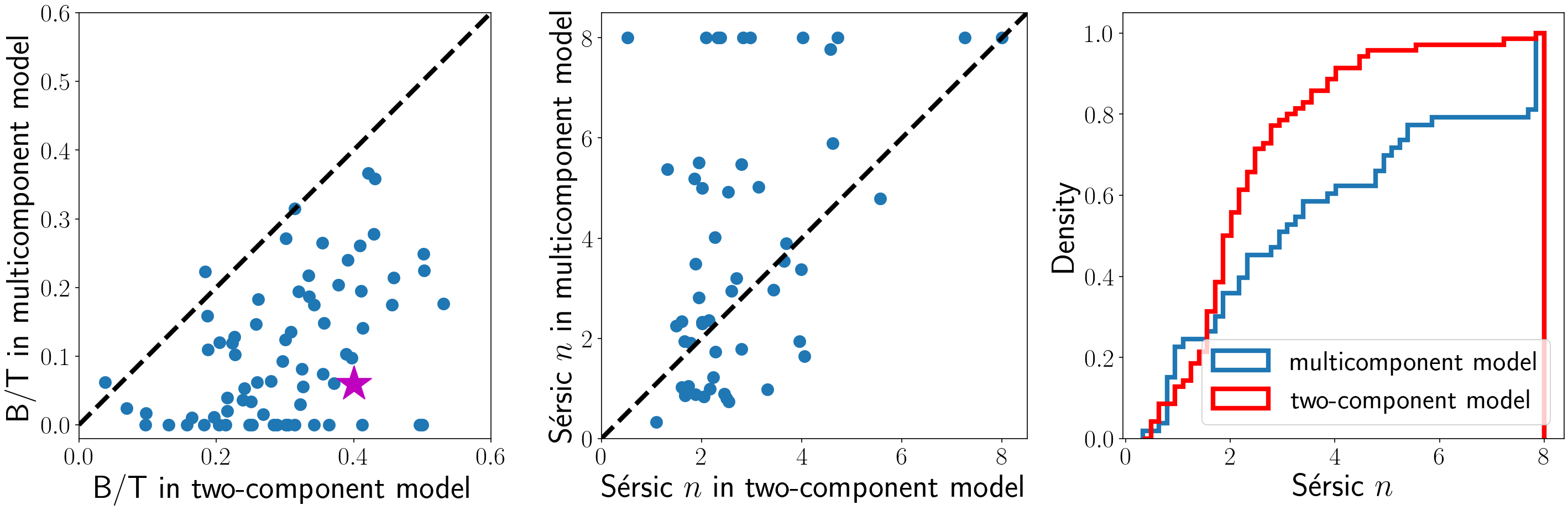}
    \caption{Comparison of bulge properties extracted from the automatic
    two-component decomposition (bulge + disk) of EGIPS galaxies and
    the multicomponent decomposition carried out in the present work.
    (\textbf{Left}): 
    Bulge-to-total contributions $B/T$.
    The magenta star marks the values reported \mbox{in \cite{2B_Kormendy}} for NGC~4565.
    (\textbf{Middle}): S\'{e}rsic indexes $n$. (\textbf{Right}):
    Cumulative distribution of S\'{e}rsic indexes.
    Note that $n=0.33$ and $n=8$ are the limits imposed for
    the S\'{e}rsic component in the decomposition.
    In the (\textbf{left}) and (\textbf{middle}) subplots,
        the dashed line indicates the one-to-one correspondence.}
    \label{fig:comp_bulges}
\end{figure}

Figure~\ref{fig:comp_bulges} presents a comparison of the
bulge-to-total ratio ($B/T$) and S\'{e}rsic indexes between two types of decomposition:
the simple two-component decomposition \mbox{(bulge + disk)} from the original EGIPS run
and our multicomponent decomposition, in which the B/PS bulge is
included as a separate component. First, the figure shows that the $B/T$ contribution typically becomes significantly smaller
in the case of multicomponent decomposition, although the decrement
varies from galaxy to galaxy. This is only natural to expect,
considering that we effectively subtract the B/PS contribution to
$B/T$ in the multicomponent decomposition.
Second, the central bulges in the multicomponent
decomposition are generally more concentrated
than those obtained in the two-component decomposition
(middle panel of  of Figure~\ref{fig:comp_bulges}).
A comparison of the cumulative distributions of S\'{e}rsic indexes
(right panel of Figure~\ref{fig:comp_bulges})
illustrates this effect clearly: for example, only about 15\%
of bulges have $n>4$ in the two-component
decomposition, whereas this fraction increases to roughly 40\% in the multicomponent case.
Overall, Figure~\ref{fig:comp_bulges} emphasizes the importance of
accounting for the composite structure of the bulges in photometric decomposition studies,
since one can entirely miss which bulge (or other central components)
the measured properties should be attributed to.

\section{Conclusions}
\label{sec:conc}

In the present work, we studied the B/PS bulges of real galaxies from
the EGIPS sample and a subsample of TNG50 galaxies with prominent B/PS bulges.
In total, we considered 71 galaxies from EGIPS and 20 galaxies from TNG50.
For the TNG50 galaxies, we prepared the images for three different bar viewing angles
(90$^\circ$, 50$^\circ$, and 30$^\circ$) using SKIRT
simulations,
adding Gaussian noise and convolving the resulting images with a PSF of 1~arcsec.

For each galaxy, we performed a 2D photometric decomposition
of the corresponding images \textcolor{black}{in the $i$-band} and obtained a
suitable photometric model (see the Supplementary Materials),
where the B/PS bulge was accounted for separately from the disk
and other bulges (if present) using our X-shaped bulge function
introduced in~\cite{Xold} and further modified in the present work.
\textcolor{black}{In essence, the function is a modified 2D S\'{e}rsic profile,
    commonly used in photometric studies, but incorporating an exponential
    density decay above the X-structure rays. The model introduces
    two additional parameters compared to the standard S\'{e}rsic function:
    the scale length of the density decay above the ray,
    and the angle $\varphi$, which defines the inclination
    of the X-structure rays relative to the disk plane.}

\textcolor{black}{Studying the measured parameters of B/PS bulges and other
components in the EGIPS galaxies and TNG50 models, we identified the following trends:}

\begin{enumerate}
\item The X-structure angles are mostly distributed between $34^\circ$ and 40$^\circ$ for
real galaxies, whereas for TNG50 galaxies the angles
span a range from about $22^\circ$ to $45^\circ$, depending on
the bar orientation with respect to the line of sight
(smaller angles correspond to bars oriented perpendicular to the LoS); see Table~\ref{tab:params}. 
The values obtained for real galaxies are consistent with the results of previous
observational and numerical studies. However, the very flat X-structures observed for
TNG50 galaxies are inconsistent with the findings of previous numerical works.
Further orbital studies of B/PS bulge in TNG50 galaxies are required to explain
how such flat X-structures are formed.
\item The X-structure angles show a significant anti-correlation with the B/PS bulge size,
that is, larger B/PS bulges tend to host more flattened X-structures (Table~\ref{tab:phi} and Figure~
\ref{fig:phi_h}).
Comparing the B/PS bulges arising from bars viewed
at different angles between the LoS and the bar major axis in the TNG50 models,
 we show that this effect is partially due to projection effects.
\item However, when examining the B/PS bulge sizes
and X-structure angles against the absolute magnitudes
and galaxy masses of real galaxies, quantities that are independent of the bar viewing angle,
we found that both larger B/PS bulges and more flattened X-structures tend
to reside in more massive galaxies (Figure~\ref{fig:ring}). This suggests that the observed
negative correlation between B/PS bulge size and
X-structure angle is not solely a projection effect
but also has a physical origin.
The trend of larger bulges residing in more massive galaxies is
consistent with previous findings that larger bars are
found in more massive galaxies. The trend of more flattened X-structures
residing in larger B/PS bulges (and, thus, in larger bars) is
a new finding of the present work. These results should be
further tested with a larger sample of galaxies with \mbox{B/PS bulges}.

\item The measured contributions of the B/PS bulge and other
bulges to total intensity are consistent with the results of previous studies
where automatic decomposition was employed.
At the same time, larger B/PS bulges are typically found in larger disks,
and generally
have larger the bulge-to-total intensity ratio, $X/T$ (Figure~\ref{fig:Xint}).
\item When studying the dependence of X-structure angles
 on disk parameters and the contribution of other bulges, $B/T$,
 we did not find any significant correlations (Table~\ref{tab:phi})

\item Comparing the B/PS bulges between EGIPS galaxies
and the TNG50 models, we find that the B/PS bulges
of real galaxies are considerably larger
(see $h_\X$ parameter values in Table~\ref{tab:params}),
both in absolute terms and relative to the disk scale.
This is consistent with previous studies of TNG50 bars,
which have shown that they are shorter than the bars of real
galaxies. At the same time, we find that
B/PS bulges in TNG50 are generally less
prominent in terms of their intensity-to-total ratio,
$X/T$, compared to real galaxies (see Table~\ref{tab:params}).
    \item Comparing the results between a simple bulge + disk decomposition and
our multicomponent decomposition,
we find that accounting for the B/PS bulge separately
is essential for obtaining realistic estimates of the bulge-to-total ratio,
$B/T$. We also found that the bulges extracted
by the two-component model tend to have smaller S\'{e}rsic
indices compared to the case where the B/PS bulge is
properly accounted for (Figure~\ref{fig:comp_bulges}).


\end{enumerate}

In general, we want to emphasize that our study is
the first to analyze such a large sample of real and model edge-on galaxies hosting
B/PS bulges using the same 2D photometric decomposition approach.
Unfortunately, B/PS bulges are objects for which
the observable properties are strongly influenced by the bar viewing angle.
However, as highlighted in this work,
their properties can still be examined in relation to global galaxy properties
that are independent of the bar viewing angle, thus providing valuable
insights into the actual vertical structure of the bars. 

\vspace{6pt} 

\supplementary{The 
 images of photometric models of all galaxies and TNG models considered in this study can be downloaded at: \linksupplementary{s1}.}

\authorcontributions{Conceptualization, Anton Smirnov, Alexander Marchuk and Natalia Sotnikova; Methodology, Anton Smirnov and Alexander Marchuk; Software, Anton Smirnov; Validation, Anton Smirnov; Formal analysis, Anton Smirnov and Alexander Marchuk; Investigation, Anton Smirnov; Resources, Anton Smirnov and Sergey Savchenko; Data curation, Anton Smirnov, Alexander Marchuk and Sergey Savchenko; Writing – original draft, Anton Smirnov; Writing – review \& editing, Anton Smirnov, Alexander Marchuk, Viktor Zozulia, Natalia Sotnikova and Sergey Savchenko; Visualization, Anton Smirnov, Alexander Marchuk and Viktor Zozulia; Supervision, Natalia Sotnikova. 
}

\funding{The authors acknowledge financial
support from the Russian Science Foundation, grant no. 24-22-00376. 
}

\dataavailability{The original contributions presented in this study are included in the article/supplementary material. Further inquiries can be directed to the corresponding author(s).
}

\acknowledgments{
This research has made use of the NASA/IPAC Extragalactic Database (NED),
which is operated by the Jet Propulsion Laboratory, California Institute of Technology,
under contract with the National Aeronautics and Space Administration.} 

\conflictsofinterest{The authors declare no conflicts of interest. 
}

\appendixtitles{yes} 
\appendixstart
\appendix

\section[\appendixname~\thesection]{Broken Disk in the Framework of 2D Decomposition}
\label{ap:disk_mess}

Since we use the broken disk model, which is not frequently applied to describe edge-on galaxies,
we provide here some clarifications on how the model works for our galaxies.
The first point to note is that it is common for the inner disk scale length to
approach infinity during the fitting $h_\mathrm{inner}\rightarrow\inf$.
In practice, it was fixed at a constant value once it exceeded the other
scale lengths by an order of magnitude. Physically, a large inner scale length implies a
plateau in the central intensity distribution of the disk.
Of course, this is a rather rough approximation of the true density distribution of the disk,
since the bar forms from the
disk’s stellar material.
However, we believe that even a plateau is more
physically
reasonable than assuming that a single exponential disk extends all the way to the center.
Therefore, we decided to keep it as is, rather than artificially constraining
the inner scale length to some small fixed value.

Secondly, because we use the broken disk model, it is not straightforward to choose
the “correct” scale length among several possible values. Therefore, in this work,
we adopt the quantity $h_\mathrm{d, 90}$ defined as
$R_\mathrm{90}/3.89$, where $R_\mathrm{90}$ is the cylindrical radius
enclosing 90\% of the disk (not total!)
intensity. For a pure exponential disk, this quantity equals the disk scale length.
For disks with multiple scale lengths, it provides an estimate of the characteristic disk size.
Since we observe the galaxies in edge-on position, $R_\mathrm{90}$ cannot be calculated
directly from the disk image due to projection effects.
Instead, we calculate it from the density profile~(Equation~(\ref{eq:disk}))
by integrating it along the cylindrical radius. $R_{90}$ is then determined
by comparing the integral value
from zero to infinity with that from zero to a specific radius $R$.

For the disk thickness, a similar issue arises.
The vertical disk profile is described by $\sech^{2/n}\left[nz/(2z_\mathrm{d})\right]$ and depends on $n$.
Therefore, the ratio $z_\mathrm{d}/h_\mathrm{d}$ does not strictly measure the disk thickness.
To study how the parameters of X-shaped bulges depend on the disk thickness,
we introduce the quantity $z_\mathrm{d,90}=z_\mathrm{90}/2.3$,
where $z_\mathrm{90}$ is defined as the height that contains 90\% of the total disk intensity.
This is measured by integrating the vertical profile with a given
$n$ value and comparing the intensity within the layer $[0, z]$ to the integral from zero to infinity.
For a pure exponential profile, $z_\mathrm{d,90}=2.3z_\mathrm{d}$.
For other profiles, the ratio $z_\mathrm{d,90}/h_\mathrm{d, 90}$ can be used as an estimate of the disk thickness.

\section[\appendixname~\thesection]{X-Shaped Bulge Model Additional Notes}
\label{ap:Xmodel}
Here, we provide a detailed explanation of why we use
the modifier $I_\mathrm{S}(x_0,y)$ in the bottom row of Equation~(\ref{eq:X}),
instead of $I_\mathrm{S}(x,y)$ as used in~\cite{Xold}.


There are two reasons for introducing this modification.
First, it prevents the model from increasing the intensity
above the rays in the vertical direction.
This was possible in the original formulation because,
while the intensity decreases as one moves from the ray toward
the center due to the introduced cutout,
the intensity of the base S\'{e}rsic profile increases.
The interaction of these two factors could lead to unrealistic solutions
in the case of slightly inclined disks if the model attempted to account
for the intensity of the \mbox{inclined disk}.

Secondly, having two counteracting factors in the intensity modifier
hurts the transparency
in the interpretation of the model parameters, since the amplitude
of the intensity dip then depends on both
$s_\X$ and the ellipticity of the isophotes of the base S\'{e}rsic profile $\varepsilon$.
In particular, in~\cite{Xold}, we suggested that the ratio
of the base S\'{e}rsic scale length to the scale length of the density truncation,
$h_{\X}/s_\X$, can be used
as a measure of the intensity dip alone.
However, in the original formulation, in case of vertical isophotes below the ray ($\varepsilon=-1$),
$h_{\X}/s_\X=1$ actually
means horizontal isophotes above the ray, since the increases and decreases compensate each other.
On the other hand, for $\varepsilon=0.4-0.5$, $h_{\X}/s_\X=1$ corresponds to
a very prominent intensity dip between the rays.
Generally, the intensity dip can be parameterized as the ratio of
intensities at the center and along the ray,
measured along the same horizontal line. For example:
\begin{equation}
    I(0, kh_\X )/I(h_\X, k h_\X) = \exp (-h_\X/s_\X)
\end{equation}
for the model defined by Equation~(\ref{eq:X}).
In the case of the original model from~\cite{Xold}, however:
\begin{equation}
    I(0, kh_\X )/I(h_\X, k h_\X) = \exp (-k/\varepsilon-h_\X/s_\X + \sqrt{1 + k^2/\varepsilon^2}),
\end{equation}
where $\varepsilon$ is the usual axis ratio $b/a$ and $k=\tan \varphi$.
In the limit $\varepsilon\rightarrow \inf$ (i.e., $b\gg a$), we have
$I(0, kh_\X )/I(h_\X, k h_\X)\rightarrow\exp(1-h_\X/s_\X$).  In other words,
when $b\gg a$, $h_\X/s_\X=1$ corresponds to the case with no density dip between the rays,
$I(0, kh_\X )/I(h_\X, k h_\X)=1$.

This was not clear in~\cite{Xold}, since the X-shaped bulges of the
simulated galaxies used to formulate the model
exhibit a clear density dip between the rays
and similar isophote ellipticity values. Now, in the updated version
described by Equation~(\ref{eq:X}),
the density dip is properly measured by the ratio $h_\mathrm{X}/s_\X$ alone.


\begin{adjustwidth}{-\extralength}{0cm}

\reftitle{References}


\PublishersNote{}
\end{adjustwidth}
\end{document}